\ifdefined\XeTeXversion\else
  \pdfoutput=1
\fi
\documentclass[11pt]{article}

\usepackage[final]{acl}

\usepackage[most]{tcolorbox}

\usepackage{times}
\usepackage{latexsym}
\usepackage{amsmath}
\usepackage{amssymb}
\usepackage{subcaption}
\usepackage{enumitem}
\usepackage{multirow}
\usepackage{booktabs}
\usepackage[T1]{fontenc}
\usepackage[utf8]{inputenc}

\usepackage{microtype}

\usepackage{inconsolata}

\usepackage{graphicx}

\title{Closing the Verification Loop: Self-Check Captioning for Long-Paragraph Detailed Audio Captioning}

\author{
  \textbf{Fengji Ma\textsuperscript{1,2\textdaggerdbl}},
  \textbf{Yan Rong\textsuperscript{1,2\textdaggerdbl}},
  \textbf{Xu Li\textsuperscript{2\textdagger}},
  \textbf{Chen Zhang\textsuperscript{2}},
  \textbf{Pengfei Wan\textsuperscript{2}},
  \textbf{Li Liu\textsuperscript{1*}}
  \\
  \textsuperscript{1}The Hong Kong University of Science and Technology (Guangzhou),
  \textsuperscript{2}Kling Team, Kuaishou Technology
  \\
  \textsuperscript{\textdagger}Project Leader
  \\
  \textsuperscript{\textdaggerdbl}Work conducted during internships at Kling Team, Kuaishou Technology.
  \\
  \textsuperscript{*}Corresponding Author:
  \href{mailto:avrillliu@hkust-gz.edu.cn}{avrillliu@hkust-gz.edu.cn}
}

\begin{document}
\maketitle
\begin{abstract}
Long-paragraph detailed audio captioning, which requires dense and
transcript-faithful descriptions of fine-grained audio content,
remains unsolved for current audio-visual multimodal language
models. We attribute this failure to two structural problems. The
first is \emph{data poverty}, as no public corpus jointly provides
long clips, paragraph captions, and verbatim-transcript fidelity.
The second is \emph{generation-mode failure}, evidenced by a
$44.8$ to $46.4$ percentage-point gap between right-audio and
shuffled-audio multiple-choice question (MCQ) accuracy. We address both within
Self-Check Captioning (SCC), a unified framework that
instantiates audio-grounded question answering as the verification
primitive at every lifecycle stage. SCC yields three artifacts.
Long-paragraph Audio Caption 50k (LACap-50k) is a $50{,}222$-clip
audio-visual corpus with $491.5$-word captions and a post-hoc
automatic speech recognition (ASR) audit.
Layer-Curvature Supervised Fine-Tuning (LC-SFT) is the first on-policy
supervised fine-tuning method
to weight tokens by intermediate-layer evidence, motivated by our
identification of Late-Layer Semantic-Entropy Collapse
(SEC). SCC-Verifier arbitrates among caption rollouts
via audio-grounded self-answering at inference. Across multiple benchmarks, our system attains state-of-the-art among
open-source captioners and is competitive with proprietary
baselines. We release LACap-50k to fill the resource gap for
long-paragraph detailed audio captioning research.
\end{abstract}

% \section{Introduction}

% \cite{Omni-Captioner}

% \section{Engines}

% ============================================================================
% Introduction.tex // Rewritten for: 2 challenges (P1 data poverty + P2
% generation-mode failure) and 3 contributions (C1 SCC framework, C2
% LACap-50k, C3 LC-SFT + SEC diagnosis).
%
% Figures: fig:gen_mode_failure (double-panel: left = perception-generation
%          gap bar chart; right = SEC token-uncertainty heatmap)
% Tables : tab:lacap_vs_corpora (dataset comparison, pulled to top of Intro)
% Cross-refs: sec:diagnostic, sec:data:stats, sec:method, sec:inference
%
% Each challenge is addressed by one or more contribution: data poverty
% by SCC's data-stage + LACap-50k; generation-mode failure by SCC's
% training + inference stages + LC-SFT.
% ============================================================================

\section{Introduction}
\label{sec:intro}

% --- Double-panel figure pulled to the very top so it appears ABOVE the
%     comparison table on the same page (float order = source order). ---------
\begin{figure*}[!t]
\centering
\begin{subfigure}[b]{0.56\textwidth}
  \centering
  \includegraphics[width=\linewidth]{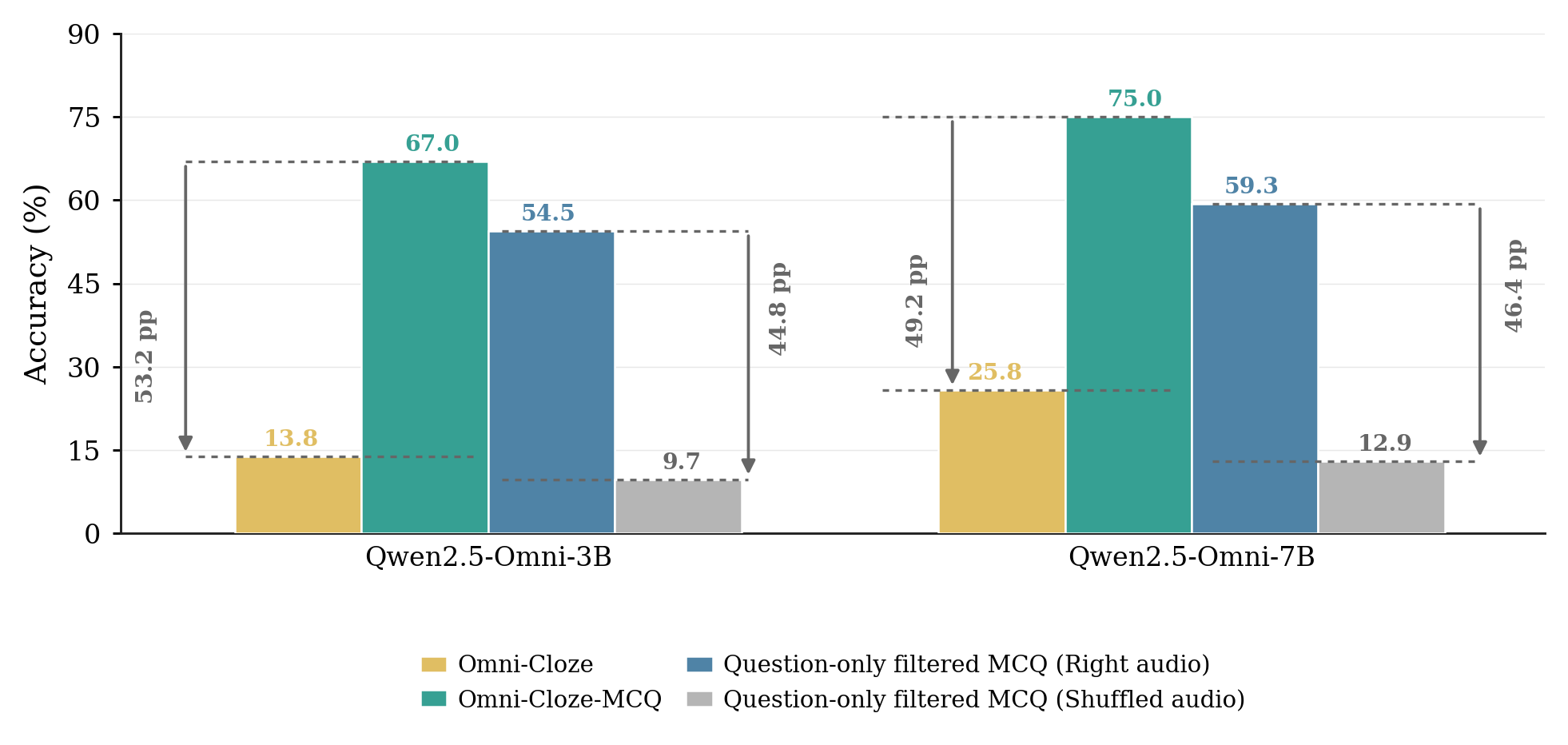}
  \caption{Omni-Cloze accuracy under four settings: cloze generation,
  multiple-choice question (MCQ) recognition, and question-only filtered
  MCQ with right vs.\
  shuffled audio. In all settings the model receives only the
  audio stream as input.}
  \label{fig:gen_mode_failure:left}
\end{subfigure}
\hfill
\begin{subfigure}[b]{0.42\textwidth}
  \centering
  \includegraphics[width=\linewidth]{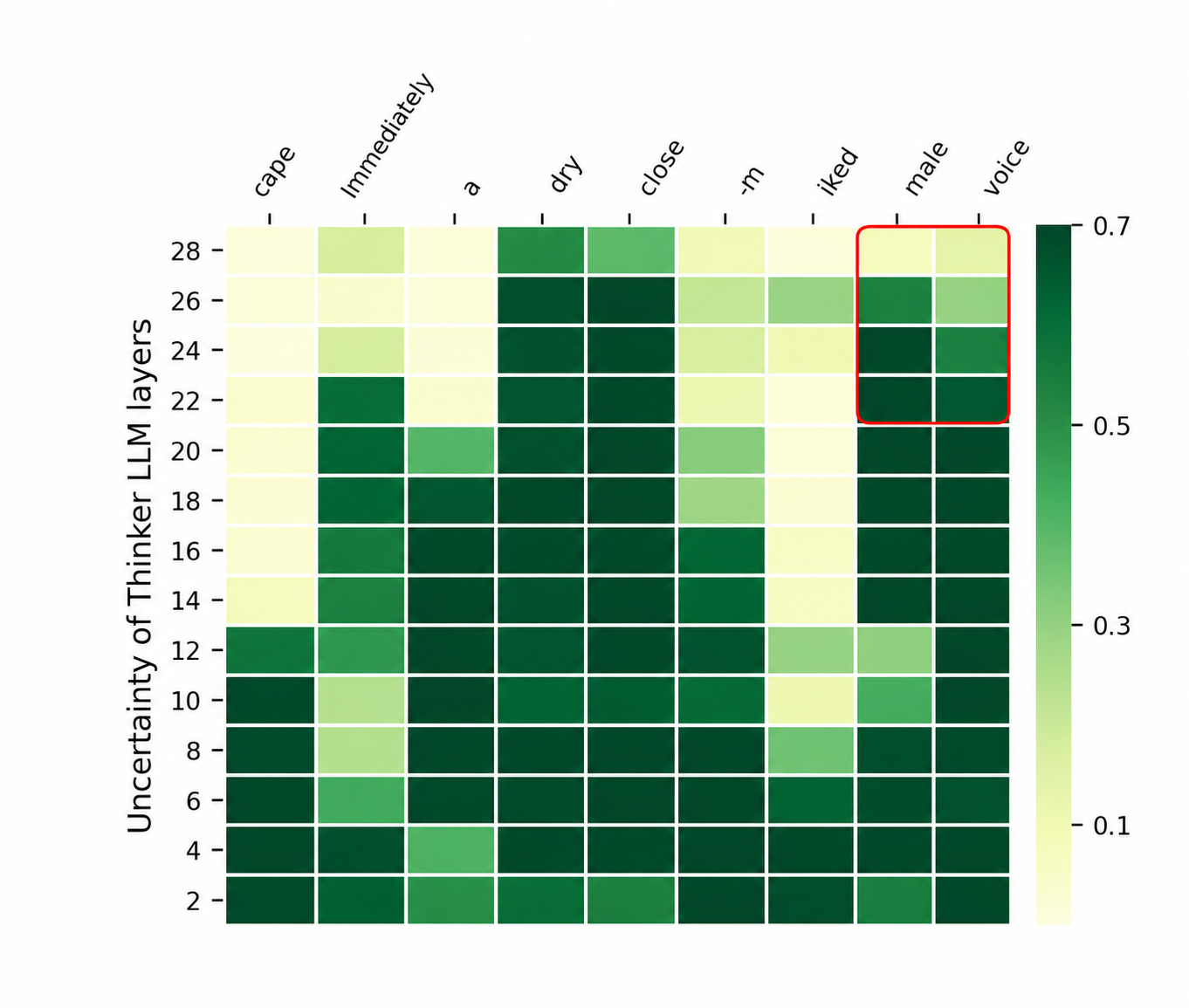}
  \caption{Token-level semantic entropy across the $28$ Thinker layers
  of Qwen2.5-Omni-7B; red box marks the hallucinated
  ``\,male voice\,'' tokens.}
  \label{fig:gen_mode_failure:right}
\end{subfigure}
\caption{\textbf{Generation-mode failure of long-paragraph audio
captioning.}
\emph{(a)} Qwen2.5-Omni-3B/7B accuracy on Omni-Cloze under four
settings: cloze generation, MCQ recognition, and the same MCQ format
restricted to a question-only filtered audio-dependent subset with
right vs.\ shuffled audio.
\emph{(b)} Token-level semantic entropy across the $28$ Thinker
layers; the red box marks tokens of the hallucinated
``\,male voice\,'' on a clip whose actual speaker is female.}
\label{fig:gen_mode_failure}
\end{figure*}

% --- Comparison-with-corpora table pulled forward so that the data-poverty
%     paragraph can reference it directly. -------------------------------------
\begin{table*}[!t]
\centering
\small
\caption{\textbf{Comparison of Long-paragraph Audio Caption 50k
(LACap-50k) with representative audio captioning corpora.}
MECAT denotes the Multi-Experts Constructed Benchmark for Fine-Grained
Audio Understanding Tasks.}
\label{tab:lacap_vs_corpora}
\setlength{\tabcolsep}{8pt}
\renewcommand{\arraystretch}{1.1}
\begin{tabular}{l r c r r}
\toprule
\textbf{Dataset}
  & \textbf{\#Clips}
  & \shortstack{\textbf{Average}\\\textbf{Duration (s)}}
  & \shortstack{\textbf{Average Text}\\\textbf{Length}}
  & \textbf{Unique Tokens} \\
\midrule
AudioCaps~\cite{AudioCaps}             & 46k             & 10.00       & 9.03    & 5.5k   \\
Clotho~\cite{Clotho}                   &  5k             & 22.50       & 11.00   & 5.5k   \\
WavCaps~\cite{WavCaps}                 & 403k            & 67.59       & 7.80    & 23.1k  \\
Auto-ACD~\cite{Auto-ACD}               & 1.5M            & 10.00       & 18.10   & 20.3k  \\
FusionAudio-1.2M~\cite{FusionAudio}    & 1.2M            & 10.00       & 47.18   & 28.1k  \\
MECAT~\cite{MECAT}                     & 40k             & 10.05   & 164.51  & 32.8k  \\
\midrule
\textbf{LACap-50k (Ours)}
                                       & 50k
                                                         & \textbf{61.12}
                                                                       & \textbf{491.51}
                                                                                 & \textbf{78.6k} \\
\bottomrule
\end{tabular}
\end{table*}

Detailed long-paragraph captioning of audio--visual clips has become a
demanding evaluation task for multimodal large language models (MLLMs).
Such captions must track sound events, report acoustic attributes,
transcribe speech verbatim, and remain faithful to the clip.
Recent audio--visual and omni-modal
MLLMs have advanced short-form audio
understanding~\citep{qwen25omni,xu2025qwen3,Salmonn,VideoSALMONN2,AudioFlamingo3},
while agent systems synthesise diverse audio types from multimodal
inputs~\citep{raudiogenie}. Yet fine-tuning on existing long-caption
corpora consistently yields outputs that fabricate sound events,
misidentify speaker attributes such as gender, and corrupt verbatim
transcripts of spoken content. We argue that this persistent failure
stems from two structural problems, neither of which has been jointly
addressed by prior work.

\paragraph{Data poverty.}
Training a paragraph-detailed long-form audio captioner requires
data that is simultaneously long in clip duration, long in caption
length, and faithful at the transcript level. No public corpus
satisfies all three (Table~\ref{tab:lacap_vs_corpora}).
Large language model (LLM)-distilled
corpora~\citep{AudioSetCaps,SoundVECaps,FusionAudio}
cap caption length below $50$ word-like tokens, long-clip
corpora~\citep{WavCaps,CASTELLA} reduce to single-sentence
captions, and MECAT~\citep{MECAT}'s $164$-word per-clip aggregate
concatenates six sentence-level sub-captions across disjoint
perspectives rather than a coherent paragraph. None cross-check
their verbatim speech transcripts against an independent automatic
speech recognition (ASR) model after construction.

\paragraph{Generation-mode failure.}
The bottleneck of long-paragraph detailed audio captioning lies in
free-form generation rather than in audio perception alone.
Pre-trained Qwen2.5-Omni-3B/7B attain $67.0\%/75.0\%$ on
Omni-Cloze~\citep{Omni-Captioner} in multiple-choice question
(MCQ) form but only $13.8\%/25.8\%$ in cloze form
(Figure~\ref{fig:gen_mode_failure:left}). On a question-only
filtered audio-dependent subset, MCQ accuracy collapses from
$54.5\%/59.3\%$ under correct audio to $9.7\%/12.9\%$ under
shuffled audio, ruling out the universal recognition-generation
format effect and text-prior shortcuts. Recent on-policy
supervised fine-tuning (SFT)
variants~\citep{DFT,ASFT,GFT,iwSFT,on-policy-sft} reweight
policy-sampled tokens using only the final-layer probability
$\pi_\theta$. Layer-by-layer probing of the language model further
reveals that hallucinated tokens hold high semantic entropy through
the middle and late layers and commit abruptly only at the final
one or two (Figure~\ref{fig:gen_mode_failure:right}), an
in-backbone signature we call \textbf{Late-Layer Semantic-Entropy
Collapse (SEC)} invisible to any reweighting from $\pi_\theta$
alone.

% \paragraph{Our approach.}
To address these two challenges jointly, we introduce
\textbf{Self-Check Captioning (SCC)}, a unified framework in which
audio-grounded question answering serves as the verification
primitive at every lifecycle stage of a long-paragraph audio
caption. At the \emph{data-construction} stage
(\S\ref{sec:data:stats}), a single Gemini model drafts a caption
from video$+$audio and re-answers its self-generated questions on
the audio stream alone (\emph{self-model} and
\emph{modality-subtraction self-check}), yielding
the \textbf{Long-paragraph Audio Caption 50k (LACap-50k)} corpus. At the
\emph{training} stage
(\S\ref{sec:method}), an external text LLM scores candidate
captions against MCQs derived from this gold. Combined with a
layer-curvature reliability signal that down-weights SEC tokens,
this constitutes \textbf{Layer-Curvature Supervised Fine-Tuning
(LC-SFT)}. At the \emph{inference} stage (\S\ref{sec:inference}),
the trained model arbitrates among caption rollouts through
audio-grounded MCQs answered from the audio alone
(\textbf{SCC-Verifier}).

% \paragraph{Contributions.}
This work makes three contributions.
% \begin{enumerate}[leftmargin=*, itemsep=2pt, topsep=2pt, label=(\arabic*)]
\begin{itemize}
\item We propose \textbf{Self-Check Captioning (SCC)}, a unified
framework that operationalises a single audio-grounded verification
primitive at the three lifecycle stages of a long-paragraph audio
caption: modality-subtraction self-check combined with self-model
self-check at the data stage, a caption-only MCQ reward reweighting
at the training stage, and an audio-grounded self-verifier
(\textbf{SCC-Verifier}) that arbitrates among multiple caption
rollouts by exploiting the captioner's MCQ-recognition strength at
the inference stage.

\item We release \textbf{LACap-50k}, the data-stage product of SCC: a
$50{,}222$-clip audio--visual corpus of paragraph-detailed captions
($491.5$ words on average) over predominantly long clips, with
sub-word vocabulary $2.4\!\times\!$ richer than the strongest prior
multi-expert corpus and post-hoc transcript-level ASR verification
absent from prior datasets. LACap-50k is released to fill the resource
gap for long-paragraph detailed audio captioning, enabling
paragraph-level captioner training and the construction of
fine-grained audio question answering (QA), long-form caption fidelity, and
modality-interaction benchmarks beyond the scope of this paper.

\item We are the first to formally diagnose long-paragraph detailed
audio captioning failure as a \textbf{generation-mode phenomenon
rather than a perception bottleneck}, evidenced by the $44.8$ to
$46.4$ percentage-point gap between right-audio and shuffled-audio
MCQ accuracy on a question-only filtered audio-dependent subset of
Omni-Cloze (Figure~\ref{fig:gen_mode_failure:left}), and we identify
\textbf{Late-Layer Semantic-Entropy Collapse (SEC)} as its
in-backbone token-level signature
(Figure~\ref{fig:gen_mode_failure:right}). Motivated by SEC, we
propose \textbf{LC-SFT}, the first on-policy SFT variant to weight
tokens by intermediate-layer evidence rather than by the
final-layer probability used in prior variants.
\end{itemize}

% ============================================================================
% Sec/RelatedWorks.tex — slim (2 subsections, citations-grouped style).
%   2.1 Long-Paragraph Detailed Captioning      — positions LACap-50k
%   2.2 On-Policy SFT and Self-Verification     — positions LC-SFT and SCC
% ============================================================================

\section{Related Work}
\label{sec:related}

\subsection{Long-Paragraph Detailed Captioning}
\label{sec:related:detailed}

Paragraph-length detailed captioning spans images~\citep{IIW},
video~\citep{Auroracap}, audio--visual
clips~\citep{VideoSALMONN2,Omni-Captioner}, time-aware dense
audio~\citep{audiomap}. The audio corpora that
condition this task span short single-sentence
captions~\citep{AudioCaps,Clotho}, large-scale LLM-distilled audio-only
corpora~\citep{WavCaps,Auto-ACD,AudioSetCaps,SoundVECaps,FusionAudio},
and recent multi-expert or long-clip
corpora~\citep{MECAT,CASTELLA,ACAVCaps}; none pairs paragraph-length
detail with predominantly long clips and post-construction
transcript-level audit, the gap LACap-50k fills
(Table~\ref{tab:lacap_vs_corpora}). Agentic audiobook systems coordinate
speech, sound effects, and music~\citep{dopamine}.
At the model side, the
Qwen-Omni~\citep{qwen25omni,xu2025qwen3},
SALMONN~\citep{Salmonn}, and Audio
Flamingo~\citep{AudioFlamingo,AudioFlamingo2,AudioFlamingo3} lines have
established strong short-form audio-visual understanding.
AudioGenie-Reasoner uses iterative evidence refinement for coarse-to-fine
audio reasoning~\citep{audiogenie_reasoner}. Our diagnostic
(Fig.~\ref{fig:gen_mode_failure}) shows that the same models
attain substantially higher audio-only MCQ accuracy than cloze accuracy
on questions derived from gold captions, indicating that the bottleneck
lies in long-form generation rather than perception alone.

\subsection{On-Policy SFT and Self-Verification}
\label{sec:related:training}

A growing line of work re-examines supervised fine-tuning through an
on-policy lens~\citep{DFT,ASFT,GFT,iwSFT,on-policy-sft}, with related
recipes pursuing full reinforcement learning under verifiable
rewards~\citep{shao2024deepseekmath}. All of these compute the token
reweighting signal from the model's final-layer probability; LC-SFT
extends the lineage by using late-layer semantic-entropy collapse, a
structural in-backbone reliability signal invisible to any final-layer
reweighting.
Iterative verification spans self-critique~\citep{self-refine},
sample aggregation~\citep{self-concistency}, evaluator
scoring~\citep{LLM-as-a-Judge}, and agentic tool calls for
captioning~\citep{Omni-Captioner}. ACE-Cap acquires missing evidence through
multi-turn audio-grounded questioning~\citep{ma2026ace}. SCC applies one audio-grounded
verification primitive across data construction, on-policy training,
and inference-time arbitration.

% % \section{Datasets}
% \input{Sec/data}

% % \section{Methodology}
% \input{Sec/method}
% \input{Sec/inference}
\section{Self-Check Captioning (SCC)}
\label{sec:scc}

% ============================================================================
% Sec/data.tex --- LACap-50k Dataset (core only).
%
% Detailed taxonomy bullets, lexical evidence, and full metric definitions
% are deferred to Sec/appendix.tex:
%   - \S\ref{app:data:taxonomy}      seven-level taxonomy with subcategories
%   - \S\ref{app:data:lexical}       lexical evidence of taxonomy coverage
%   - \S\ref{app:data:asr}           ASR metric definitions and per-language
%   - \S\ref{app:data:non-speech}    post-hoc non-speech grounding protocol
% ============================================================================

% ─── Combined LACap-50k overview figure (double-column, 3+2 layout) ─────────
\begin{figure*}[t]
\centering

% Row 1: duration donut + 2 hot-word clouds (all at thumbnail size)
\begin{subfigure}[t]{0.22\textwidth}
  \centering
  \includegraphics[width=\linewidth]{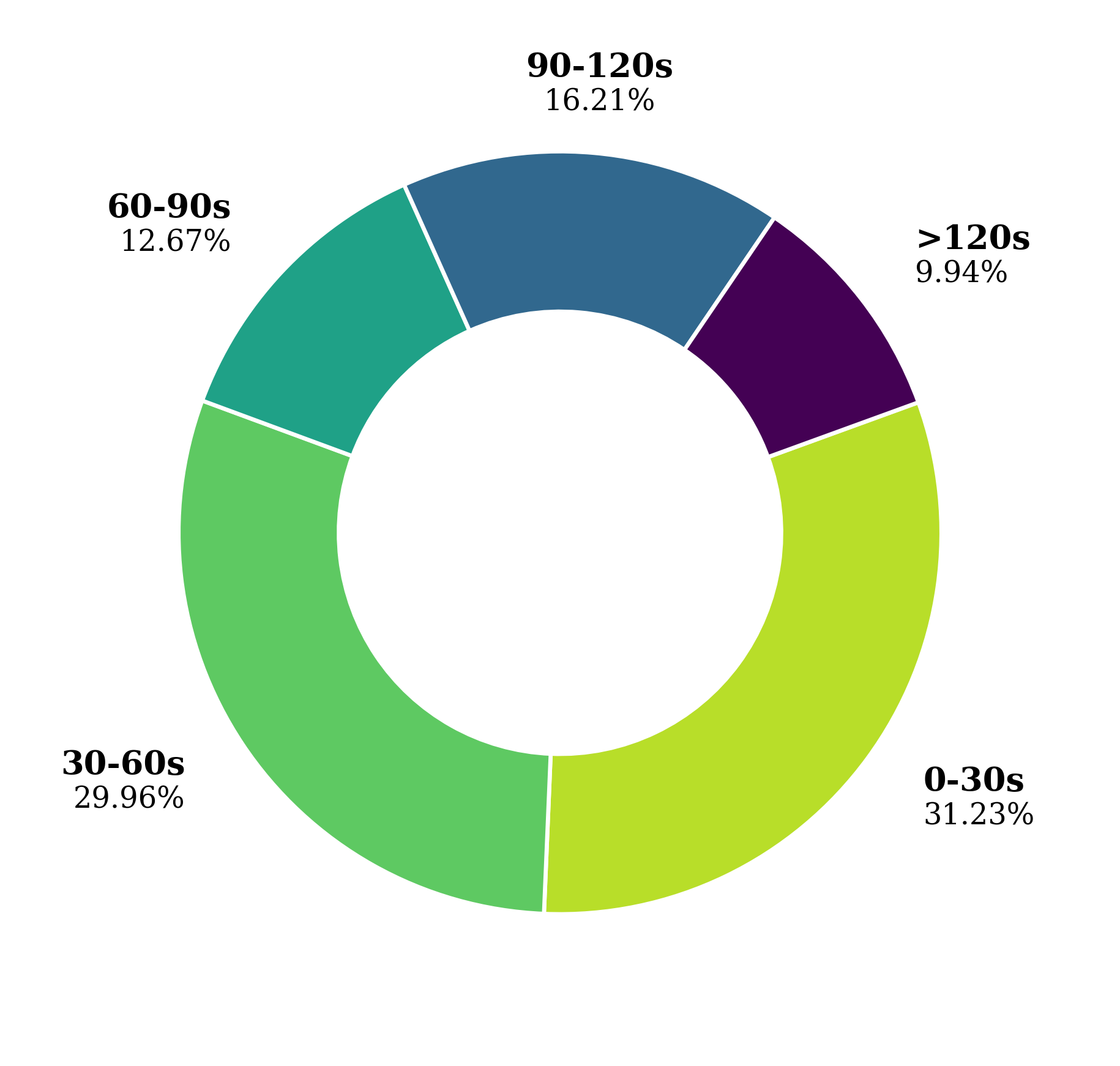}
  \caption{Clip duration distribution across five buckets.}
  \label{fig:lacap:duration}
\end{subfigure}
\hfill
\begin{subfigure}[t]{0.37\textwidth}
  \centering
  \includegraphics[width=\linewidth]{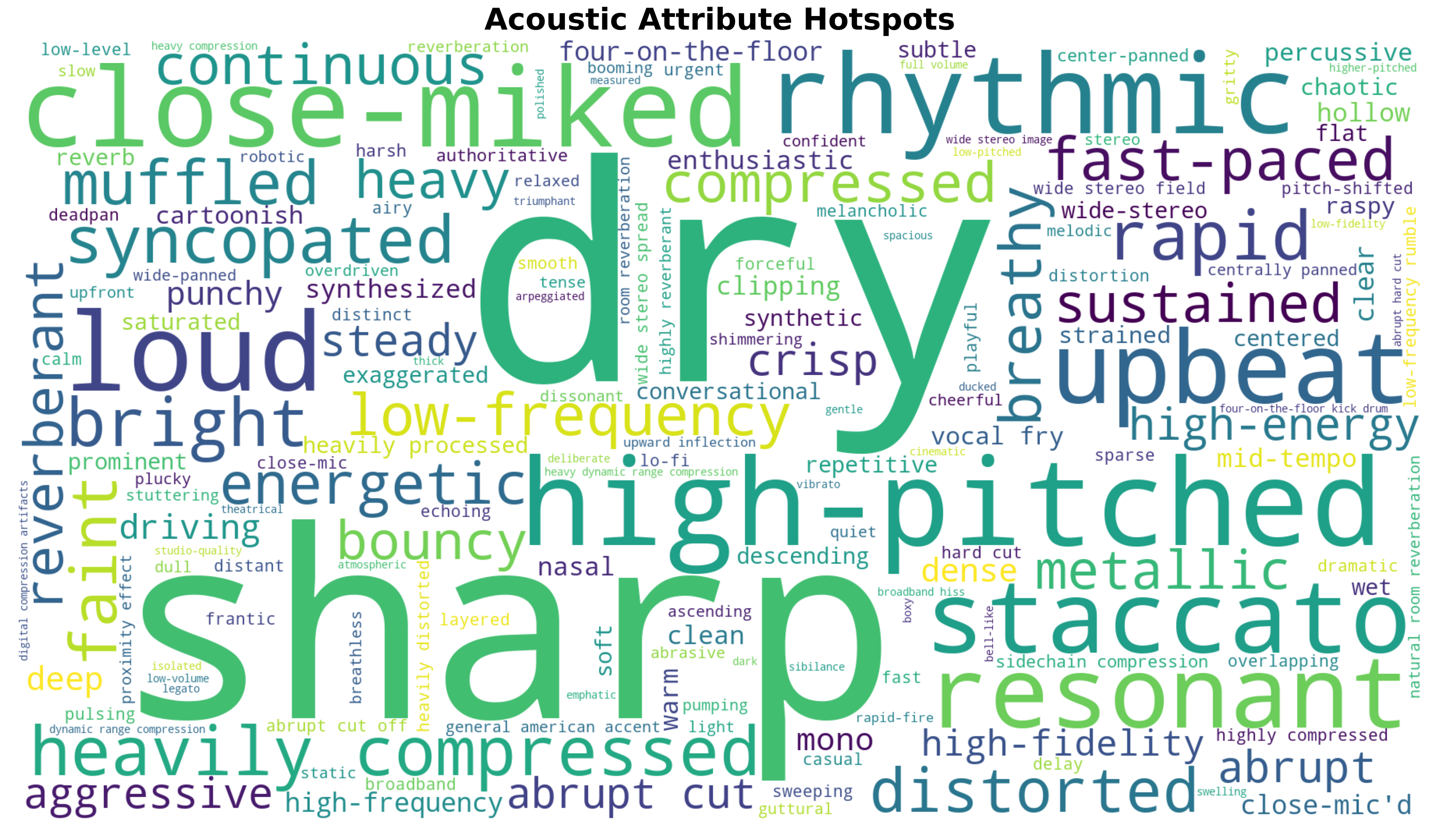}
  \caption{Acoustic-attribute hot words.}
  \label{fig:lacap:wc_acoustic}
\end{subfigure}
\hfill
\begin{subfigure}[t]{0.37\textwidth}
  \centering
  \includegraphics[width=\linewidth]{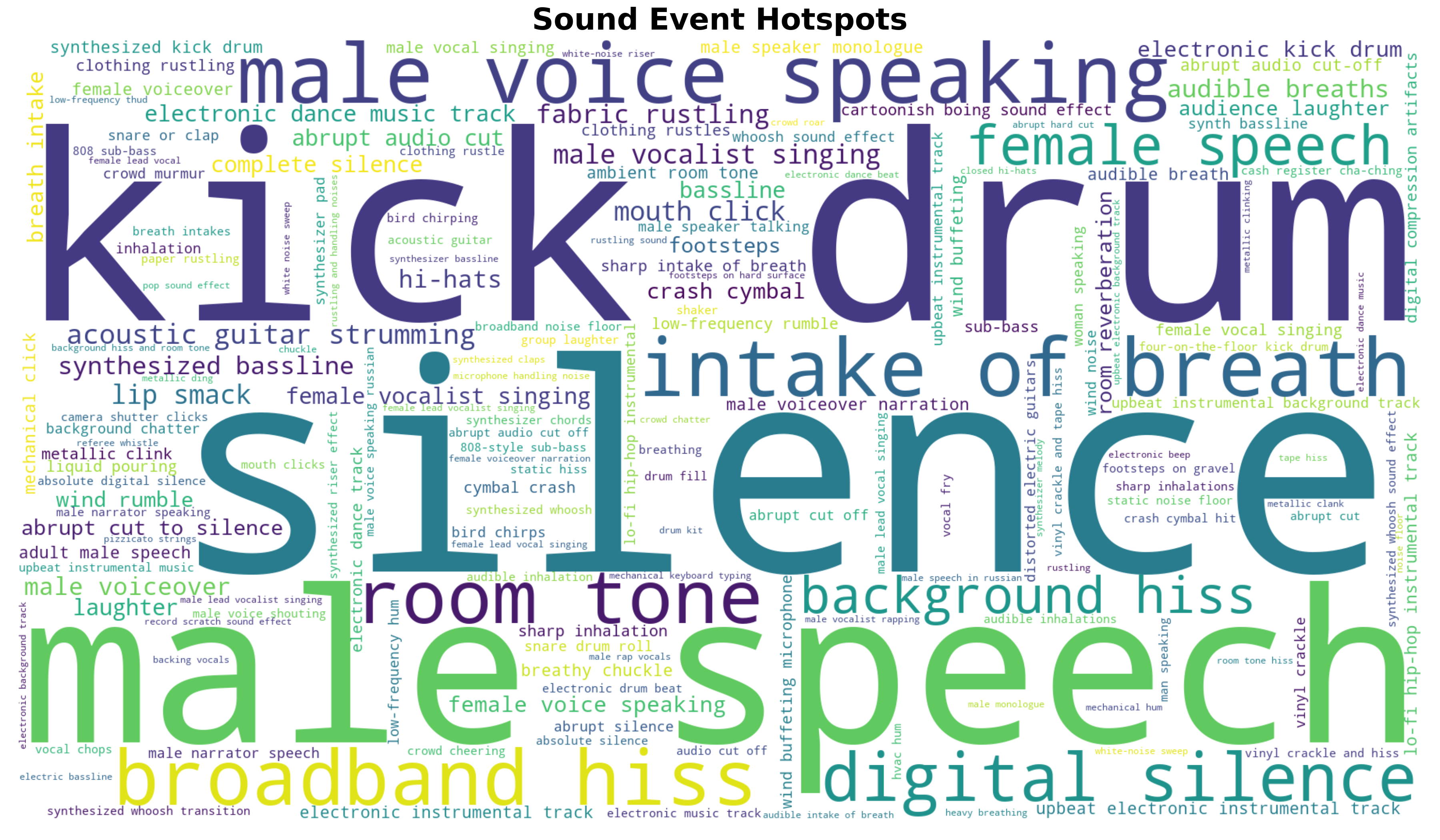}
  \caption{Sound-event hot words.}
  \label{fig:lacap:wc_events}
\end{subfigure}

\vspace{0.6em}

% Row 2: enlarged taxonomy donut + caption length KDE
\begin{subfigure}[t]{0.43\textwidth}
  \centering
  \includegraphics[width=\linewidth]{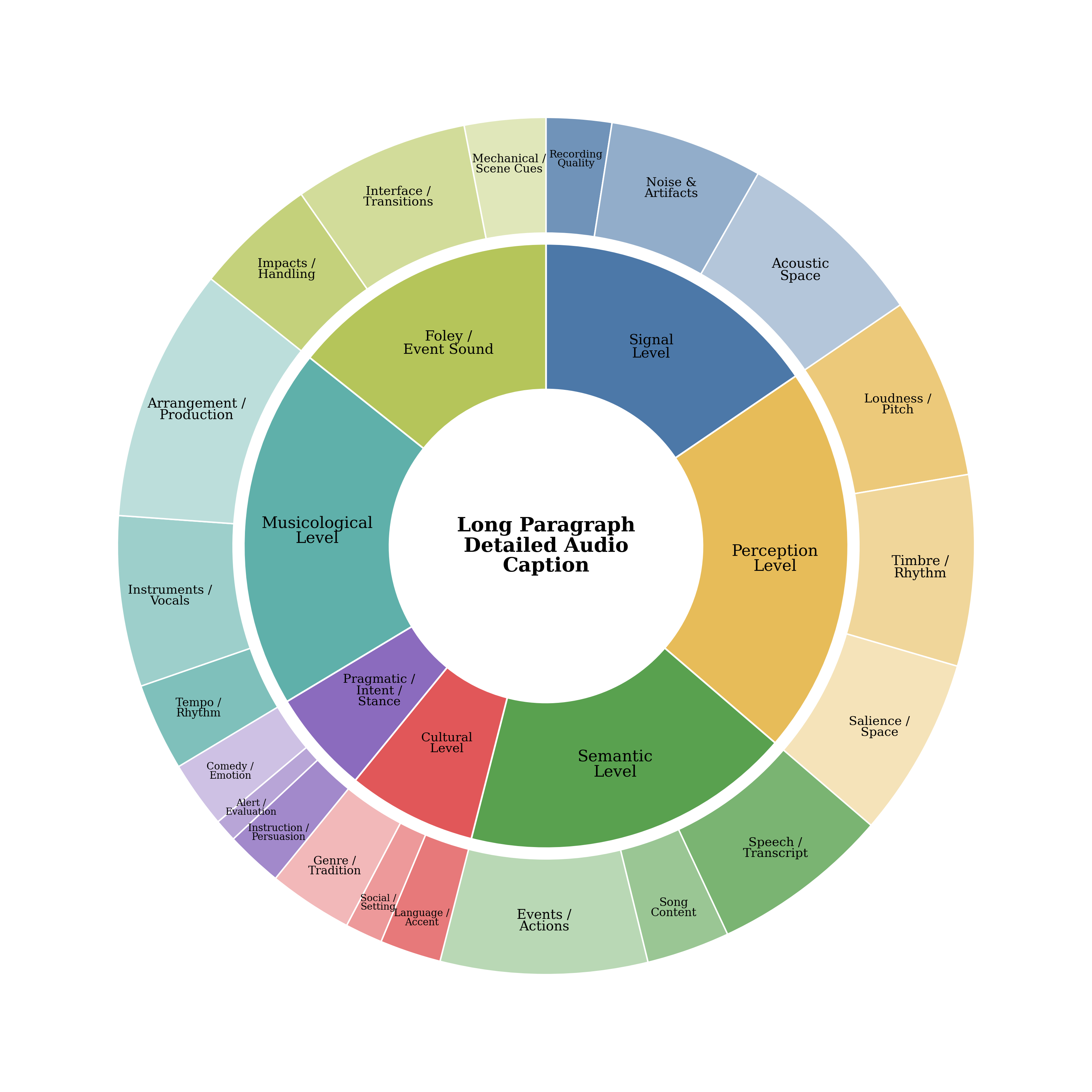}
  \caption{Seven-level audio analysis taxonomy with three
  subcategories per level.}
  \label{fig:lacap:taxonomy}
\end{subfigure}
\hfill
\begin{subfigure}[t]{0.55\textwidth}
  \centering
  \includegraphics[width=\linewidth]{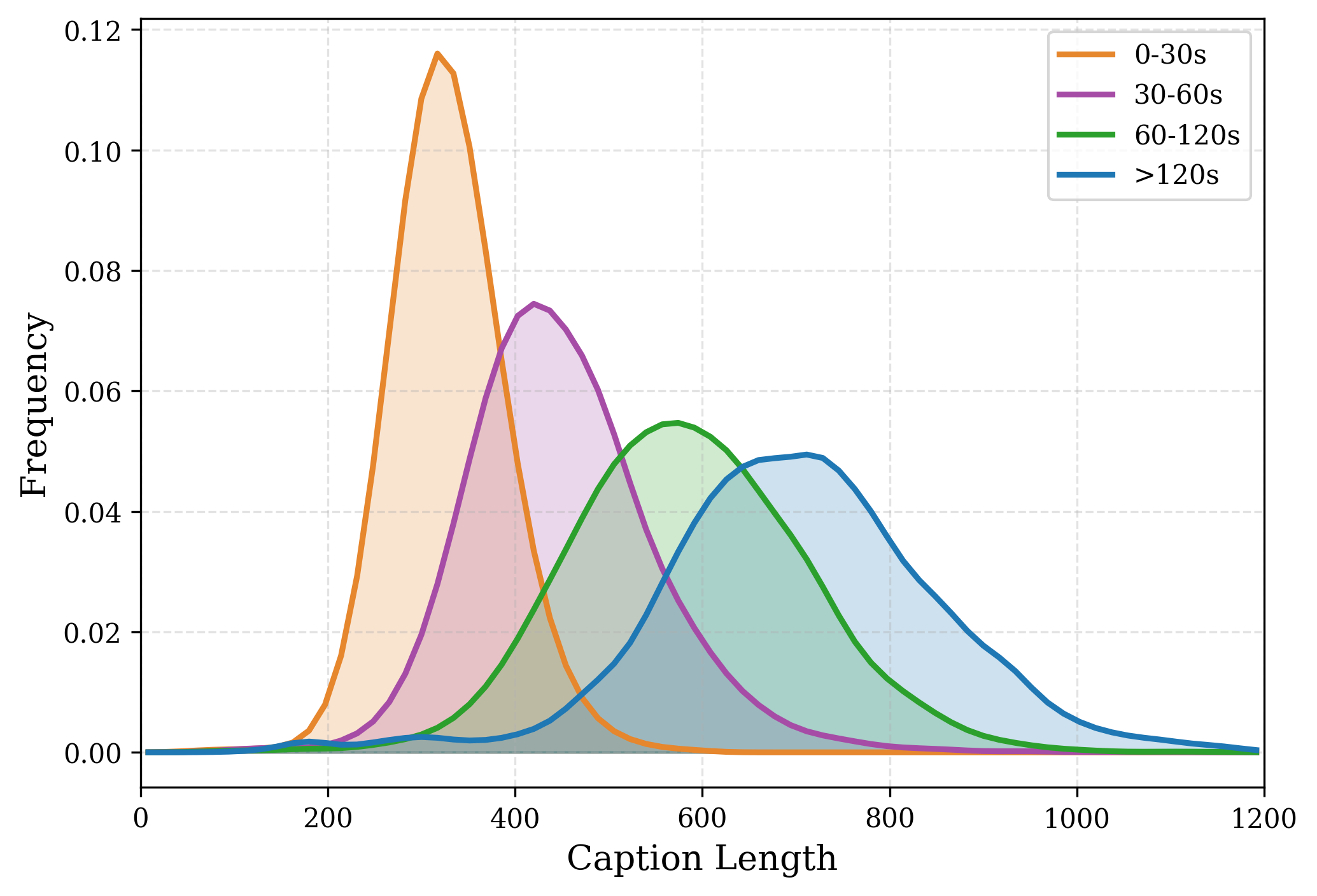}
  \caption{Per-bucket caption length, word-like units.}
  \label{fig:lacap:caplen}
\end{subfigure}

\caption{\textbf{LACap-50k overview.}
\emph{Top}: clip-duration distribution and hot-word clouds for
acoustic attributes and sound events.
\emph{Bottom}: seven-level audio analysis taxonomy and per-bucket
caption-length kernel density.}
\label{fig:lacap_overview}
\end{figure*}

\subsection{Data Stage: LACap-50k Construction}
\label{sec:data}

LACap-50k is the data-stage product of Self-Check Captioning: a
$50{,}222$-clip audio--visual corpus paired with one long-paragraph
caption per clip ($491.5$ word-like tokens on average, $78.6$\,k
unique sub-word tokens). We sample the $50{,}222$ source videos from
ASID-1M~\citep{ASID1M} based on audio type, duration, and sound-event coverage;
Figure~\ref{fig:lacap_overview} shows their distribution. The entire construction loop uses
Gemini3.1-Pro for both caption generation and verification. We term
this design \emph{self-model self-check}: the same model that authors the
initial caption from the audio--visual clip also generates the
verification questions and supplies their answers, removing the
inter-model disagreement noise that a multi-model pipeline would
introduce. Crucially, the verification questions are open-form
queries targeting the fine-grained acoustic details of specific
audio events, and they are re-answered on the audio stream alone,
with the video and the initial caption deliberately withheld. We
call this audio-only re-listening pass \emph{modality-subtraction
self-check}: it strips away visual evidence so that the revised
caption is anchored in audio content rather than in details
inferred from the video, mitigating the modality bias that
audio--visual inputs would otherwise leave in the draft. Captions
span the seven-level audio analysis taxonomy of
Figure~\ref{fig:lacap:taxonomy}, covering signal, perception,
semantic, cultural, pragmatic, musicological, and Foley event-sound
facets. Subcategory definitions and lexical evidence are deferred
to Appendices~\ref{app:data:taxonomy}
and~\ref{app:data:lexical}.

\paragraph{Distribution.}
LACap-50k deliberately spans short, medium, and long clips: $31.23\%$
in $0$--$30$\,s, $29.96\%$ in $30$--$60$\,s, and the remaining
$38.81\%$ beyond $60$\,s (Figure~\ref{fig:lacap:duration}). Per-bucket
caption length grows monotonically with clip duration
(Figure~\ref{fig:lacap:caplen}), the mode shifting from $\sim\!320$
tokens for $0$--$30$\,s clips to $\sim\!700$ tokens beyond $120$\,s.

\paragraph{Transcript fidelity.}
\label{sec:data:stats}
We cross-check the verbatim speech segments inside each caption
against the open-source Qwen3-ASR-1.7B model~\citep{Qwen3-ASR} after
construction. Over $10{,}975$ speech-heavy captions yielding $36{,}809$
transcript spans, LACap-50k attains a micro-averaged character error
rate (CER) of $6.23\%$ and word error rate (WER) of $9.21\%$
(Table~\ref{tab:asr_fidelity}); the English subset is markedly
cleaner than the non-English subset, consistent with the relative
coverage of the reference ASR's training data.
Metric definitions and threshold-share statistics are deferred to
Appendix~\ref{app:data:asr}.

\begin{table}[t]
\centering
\scriptsize
\setlength{\tabcolsep}{2.2pt}
\begin{tabular}{l c c c c}
\toprule
\textbf{Subset}
  & \shortstack{\textbf{Micro}\\\textbf{CER}}
  & \shortstack{\textbf{Macro}\\\textbf{CER}}
  & \shortstack{\textbf{Micro}\\\textbf{WER}}
  & \shortstack{\textbf{Macro}\\\textbf{WER}} \\
\midrule
English      & 5.58  & 7.41  & 8.60  & 10.86 \\
Non-English  & 9.31  & 11.62 & 10.20 & 13.88 \\
\midrule
\textbf{All} & \textbf{6.23} & \textbf{8.32} & \textbf{9.21} & \textbf{11.51} \\
\bottomrule
\end{tabular}
\caption{\textbf{Transcript fidelity of LACap-50k} (\%) on
$36{,}809$ transcript spans from $10{,}975$ speech-heavy captions,
against the open-source Qwen3-ASR-1.7B
reference~\citep{Qwen3-ASR}.}
\label{tab:asr_fidelity}
\end{table}

\paragraph{Post-hoc non-speech event grounding.}
\label{sec:data:non-speech-grounding}
To complement the transcript-fidelity audit, we evaluate non-speech event
grounding on $1{,}000$ final LACap-50k captions: $500$ clips of $0$--$30$ s
and $500$ of $30$--$60$ s.
Qwen3.6-27B~\citep{qwen3.6-27b} extracts explicit atomic event claims from
each caption. We canonicalise, filter, and deduplicate these claims using
the eligibility rules before audio verification.

The open-source Frame-wise Language-Audio Modeling system
OpenFLAM~\citep{OpenFLAM} receives the source audio and each event query. Of
$3{,}985$ eligible claims,
$3{,}770$ ($94.60\%$) are corroborated, $151$ ($3.79\%$) are not corroborated,
and $64$ ($1.61\%$) are uncertain. Appendix~\ref{app:data:non-speech}
provides the complete protocol and reporting definitions.

\paragraph{Positioning and release.}
A head-to-head comparison against representative audio captioning
corpora is reported in Table~\ref{tab:lacap_vs_corpora}
(\S\ref{sec:intro}): LACap-50k is roughly an order of magnitude longer
in average caption length than the strongest prior LLM-distilled
corpus, three times longer than the strongest prior multi-expert
corpus, and the only listed corpus paired with a post-construction
transcript-level ASR cross-check. LACap-50k is released under a
permissive license to fill the resource gap for long-paragraph
detailed audio captioning research; together with the paired
multimodal initial draft and audio-only revised final caption, and
the audio-only verification question--answer pairs from the SCC
self-check loop, the release enables paragraph-level captioner
training and the construction of fine-grained audio QA, long-form
caption fidelity, and modality-interaction benchmarks beyond the
scope of this paper.

% ============================================================================
% Sec/method.tex --- LC-SFT methodology (compressed, core only).
%
% Detailed derivations and justifications moved to Sec/appendix.tex:
%   - \S\ref{app:method:curv-design}  why second-order curvature
%   - \S\ref{app:method:gradient}     full gradient expression + interp
%   - \S\ref{app:method:two-stage}    two-stage training details
%
% Main-text structure (kept):
%   4.1 Setting                       (notation + two-stage framing)
%   4.2 Response Group and Reward     (eq:group, eq:reward, eq:adv)
%   4.3 Layer-Curvature Signal        (eq:earlyexit, eq:curv, eq:wt)
%   4.4 The LC-SFT Objective          (eq:percand, eq:groupobj)
% ============================================================================

\subsection{Training Stage: Layer-Curvature SFT (LC-SFT)}
\label{sec:method}

We instantiate the on-policy SFT scheme as \textbf{LC-SFT}, an
on-policy fine-tuning objective in which a verifiable caption-only
reward selects which candidate captions drive the gradient, and a
layer-curvature reliability signal computed from the policy's
intermediate representations selects which tokens within each
candidate carry that gradient.

\paragraph{Setting and two-stage training.}
\label{sec:method:setting}
Let
$\pi_\theta(y\!\mid\!x)\!=\!\prod_{t=1}^{T}\!\pi_\theta(y_t\!\mid\!x,y_{<t})$
denote the autoregressive policy of the audio-visual language model.
We train within the on-policy SFT
regime of \citet{on-policy-sft,iwSFT} in two stages. Stage~1 applies
standard SFT on $\mathcal{D}\!=\!\{(x,y^{\star})\}$ to grant
paragraph-level capacity. Stage~2 (LC-SFT, defined below) refines
the Stage-1 checkpoint by addressing residual generation-mode
failures visible only in the SFT-tuned policy's own rollouts, from
which all Stage-2 response groups (\S\ref{sec:method:reward}) are
sampled. Full hyperparameters appear in
Appendix~\ref{app:method:two-stage}.

\subsubsection{Response Group and Verifiable Reward}
\label{sec:method:reward}

For each input $x$ we construct a response group of $K$ caption
candidates,
\begin{equation}
\mathcal{G}_x \;=\; \bigl\{\,y_{\mathrm{gold}},\;y_{r_1},\;\dots,\;y_{r_{K-1}}\,\bigr\},
\label{eq:group}
\end{equation}
where $y_{\mathrm{gold}}$ is the LACap-50k reference for $x$
(\S\ref{sec:data:stats}) and $\{y_{r_i}\}_{i=1}^{K-1}$ are $K\!-\!1$
multinomial samples freshly drawn from the current policy
$\pi_\theta(\cdot\!\mid\!x)$ at every gradient step. The choice of
$K$ is reported in \S\ref{sec:experiments}.

Each candidate $y_k\!\in\!\mathcal{G}_x$ is scored by a caption-only
MCQ reward. Offline, a frozen external text language model converts
$y_{\mathrm{gold}}$ into a pool $\mathcal{Q}_x$ of cloze-style
questions about fine-grained audio facts in the clip, each paired
with four close distractors and a ``not given'' option. At training
time, the same evaluator answers $\mathcal{Q}_x$ conditioned only on
the candidate text $y_k$ (audio and video removed), and we record
the accuracy
\begin{equation}
R(y_k) \;=\;
\frac{1}{|\mathcal{Q}_x|}
\sum_{q\in\mathcal{Q}_x}
\mathbf{1}\!\bigl[\mathrm{eval}(y_k,q)=\mathrm{answer}(q)\bigr].
\label{eq:reward}
\end{equation}
The caption-only protocol forces $y_k$ to retain every fact that
$\mathcal{Q}_x$ probes, so $R(y_k)$ measures retention of
$y_{\mathrm{gold}}$'s verified audio-fact content rather than
stylistic similarity (a comparison with the inference-time verifier
identity appears in
Appendix~\ref{app:inference:relation-to-training}). We standardise
$R$ within each group to obtain the per-candidate advantage
\begin{equation}
A(y_k) \;=\;
\frac{R(y_k)-\mu(\mathcal{G}_x)}{\sigma_R(\mathcal{G}_x)+\epsilon},
\qquad k=1,\dots,K,
\label{eq:adv}
\end{equation}
where $\mu(\mathcal{G}_x)$ and $\sigma_R(\mathcal{G}_x)$ are the
empirical mean and standard deviation of $R$ over $\mathcal{G}_x$
and $\epsilon\!>\!0$ a small constant. Because $\mathcal{Q}_x$ is
derived from $y_{\mathrm{gold}}$,
$R(y_{\mathrm{gold}})\!\approx\!1$ by construction and
$A(y_{\mathrm{gold}})$ is the largest value in the group almost
surely, serving as a supervised anchor. Below-mean candidates
receive $A(y_k)\!<\!0$ and contribute an unlikelihood gradient.

\subsubsection{Layer-Curvature Reliability Signal}
\label{sec:method:curv}

A scalar score $A(y_k)$ cannot distinguish reliably-learned tokens
from individual hallucinated tokens inside an otherwise high-reward
candidate. We obtain a per-token reliability signal directly from
the policy's own intermediate representations through a
parameter-free early-exit probe that reuses the model's final
layer-norm $\mathrm{Norm}(\cdot)$ and language-model head
$W_{\mathrm{lm}}$ without modification.

For each assistant token $y_{k,t}$ and each layer
$\ell\!\in\!\mathcal{L}\!=\!(l_1,\dots,l_m)$ from a fixed set of
late Thinker layers, the early-exit predictive distribution and the
per-layer negative log-likelihood of the realised token are
\begin{align}
p^{(\ell)}_{k,t} \;&=\;
\mathrm{softmax}\!\bigl(W_{\mathrm{lm}}\,\mathrm{Norm}(h^{(\ell)}_{k,t})\bigr),\nonumber\\
U^{(\ell)}_{k,t} \;&=\; -\log p^{(\ell)}_{k,t}\bigl(y_{k,t}\bigr),
\label{eq:earlyexit}
\end{align}
where $h^{(\ell)}_{k,t}$ is the hidden state at layer $\ell$. The
trajectory $(U^{(l_1)}_{k,t},\dots,U^{(l_m)}_{k,t})$ describes how
the model resolves its prediction for $y_{k,t}$ across depth.
For computational tractability, we use realised-token negative
log-likelihood (NLL) $U$
(surprisal) as a proxy for token-level semantic entropy rather than
computing entropy over the full predictive distribution; the SEC name
refers to the resulting late-layer collapse pattern.

\paragraph{SEC and its detection.}
On audio-faithful tokens this trajectory descends smoothly across
the late layers. On hallucinated tokens it remains elevated through
the middle and late layers and drops abruptly only at the final one
or two layers (Late-Layer Semantic-Entropy Collapse, SEC;
Fig.~\ref{fig:gen_mode_failure:right}). Because the final-layer
probability $\pi_\theta(y_{k,t}\!\mid\!\cdot)$ is unchanged by
whether the preceding trajectory was smooth or abrupt, it is
structurally blind to SEC, and capturing SEC requires a feature
that compares predictions across depth. We define the
\emph{layer-curvature} of token $(k,t)$ as the smoothed second-order
difference of $U^{(\ell)}_{k,t}$ summed over the intermediate layer
indices,
\begin{equation}
\mathrm{Curv}_{k,t} \;=\;
\sum_{j=2}^{m-1}
\sqrt{\!\bigl(\Delta U^{(j+1)}_{k,t}-\Delta U^{(j)}_{k,t}\bigr)^2+\epsilon^2},
\label{eq:curv}
\end{equation}
where
$\Delta U^{(j)}_{k,t}\!=\!U^{(l_j)}_{k,t}\!-\!U^{(l_{j-1})}_{k,t}$
and the smoothed envelope $\sqrt{u^2+\epsilon^2}$ approximates
$|u|$ differentiably. The second-order form is essential.
$\mathrm{Curv}_{k,t}$ vanishes on any constant-slope trajectory and
is large only on trajectories with a true kink, isolating the SEC
signature. A formal justification for this choice appears in
Appendix~\ref{app:method:curv-design}.

We convert the curvature to a multiplicative stop-gradient token
weight,
\begin{equation}
w_{k,t} \;=\;
\mathrm{sg}\!\Bigl(\exp\!\bigl(-\kappa\,\mathrm{Curv}_{k,t}\bigr)\Bigr)
\;\in\;(0,1],
\label{eq:wt}
\end{equation}
where $\mathrm{sg}[\cdot]$ is the stop-gradient operator that
prevents the model from masking the kink by reshaping intermediate
representations, and $\kappa\!>\!0$ is a validation-selected
suppression strength (value in \S\ref{sec:experiments}).

\subsubsection{The LC-SFT Objective}
\label{sec:method:obj}

Combining $A(y_k)$ from Eq.~\eqref{eq:adv} and $w_{k,t}$
from Eq.~\eqref{eq:wt}, the per-candidate term is the $w$-weighted
average of the per-token log-probability
$\ell_{k,t}\!:=\!\log\pi_\theta(y_{k,t}\!\mid\!x,y_{k,<t})$,
\begin{equation}
\widetilde{\mathcal{L}}(y_k;\,x)
\;=\;
-\,\frac{\sum_{t=1}^{T_k}\,w_{k,t}\,\ell_{k,t}}
        {\sum_{t=1}^{T_k}\,w_{k,t}+\epsilon},
\label{eq:percand}
\end{equation}
and the group-level loss is the $A$-weighted linear combination
\begin{equation}
\mathcal{L}(\mathcal{G}_x)
\;=\;
\sum_{k=1}^{K}\,A(y_k)\,\widetilde{\mathcal{L}}(y_k;\,x),
\label{eq:groupobj}
\end{equation}
yielding the training objective
$\mathcal{L}_{\mathrm{LC}}(\theta)\!=\!\mathbb{E}_{x\sim\mathcal{D}}
[\mathcal{L}(\mathcal{G}_x)]$, approximated by minibatch averaging
over freshly sampled rollouts at every step. The advantage
$A(y_k)$ thus selects which candidates contribute to the gradient
(positive as maximum-likelihood, negative as unlikelihood) while
$w_{k,t}$ selects which tokens within each candidate carry it. The
full gradient and its two-axis interpretation appear in
Appendix~\ref{app:method:gradient}.

% ============================================================================
% Sec/inference.tex --- SCC-Verifier inference-stage self-check.
% Methodology only (no quantitative numbers); M, N, T are deferred to
% \S\ref{sec:experiments}.
%
% Structure:
%   §?.1 Protocol         (3 steps: candidate generation, MCQ generation, arbitration)
%   §?.2 Relation to the Training Reward
%   §?.3 Cost
%
% Cross-refs:
%   - sec:method, sec:method:reward, sec:method:curv (method.tex)
%   - sec:experiments                                 (Experiments)
%   - sec:intro                                       (Introduction.tex)
% No new bib citations introduced.
% ============================================================================

\subsection{Inference Stage: SCC-Verifier}
\label{sec:inference}

A single inference pass remains exposed to sampling stochasticity
even after LC-SFT (\S\ref{sec:method}). We close the SCC loop at
inference with \textbf{SCC-Verifier}, which re-uses the
LC-SFT-trained model itself as an audio-grounded multiple-choice
answerer to arbitrate among candidate captions in three steps.
\textit{(i)} The policy $\pi_\theta$ produces $M$ candidates for an
audio-visual clip $x$, one greedy decoding $y^{(0)}$ and $M\!-\!1$
multinomial samples $y^{(1)},\dots,y^{(M-1)}$ at temperature $T$.
\textit{(ii)} For each $y^{(m)}$, a frozen external text language
model extracts $N$ atomic audio-grounded claims and converts each
into a five-option MCQ whose correct option is the claim asserted
by $y^{(m)}$ and whose distractors are minimally perturbed
alternatives along the same audio facet plus a ``not given''
option, yielding a pool $\mathcal{Q}^{(m)}$ that encodes precisely
what $y^{(m)}$ claims about the audio. \textit{(iii)} For each
$(y^{(m)},q)$ with $q\!\in\!\mathcal{Q}^{(m)}$, the LC-SFT-trained
model answers $q$ in audio-only MCQ mode on $x_{\mathrm{audio}}$
with $y^{(m)}$ withheld, and the candidate score is the fraction
of its own claims that survive,
\begin{equation}
S(y^{(m)}) = \frac{1}{|\mathcal{Q}^{(m)}|} \sum_{q\in\mathcal{Q}^{(m)}} \mathbf{1}\bigl[a_\theta(x_{\mathrm{audio}}, q) = a^{(m)}_q\bigr],
\label{eq:cts:score}
\end{equation}
where $a_\theta$ denotes the model's audio-only answer and
$a^{(m)}_q$ the option asserted by $y^{(m)}$. SCC-Verifier outputs
$y^{\star}\!=\!\arg\max_m S(y^{(m)})$, complementary by
construction to the training-time MCQ reward of
\S\ref{sec:method:reward}
(Appendix~\ref{app:inference:relation-to-training}).

\section{Experiments}
\label{sec:experiments}

This section reports the empirical performance of LC-SFT
(\S\ref{sec:method}) and the SCC-Verifier inference protocol
(\S\ref{sec:inference}) on four benchmarks that jointly probe
audio understanding through downstream question answering and
direct detailed-caption fidelity: \textbf{Massive Multi-Task Audio
Understanding and Reasoning (MMAU)}~\citep{mmau},
\textbf{MMAR (Deep Reasoning in Speech, Audio, Music, and Their
Mix)}~\citep{mmar}, \textbf{Massive Multi-Task Spoken Language Understanding
and Reasoning (MMSU)}~\citep{mmsu}, and
\textbf{Omni-Cloze}~\citep{Omni-Captioner}. The first three follow
a \emph{caption-as-evidence} protocol in which the generated
caption is the sole audio-derived evidence supplied to a frozen
text-only judge language model (LLM-as-a-judge), which then answers
the benchmark questions; the fourth scores the captions directly
via cloze-style multiple-choice blanks. Higher scores indicate greater caption
fidelity; Appendix~\ref{sec:experiments:setup} details the setup. All Gemini
results use Gemini3.1-Pro-Preview~\citep{gemini}, denoted
\mbox{\textbf{Gemini3.1-Pro}} throughout.

% \subsection{Setup}
% \label{sec:experiments:setup}

% ----------------------------------------------------------------------------
% PLACEHOLDER --- fill once training/inference settings are finalized.
% Suggested content:
%   - Backbone: Qwen2.5-Omni-7B
%   - Two-stage training: Stage 1 standard SFT on LACap-50k; Stage 2 LC-SFT
%     with K=??, kappa=??, optimizer, learning rate, # GPUs, # steps
%   - SCC-Verifier inference: M=??, N=??, T=??
%   - Text-only judge LLM (LLM-as-a-judge backbone): GPT-4o or other
%   - Hardware
% ----------------------------------------------------------------------------

% \textit{(Setup details on training hyperparameters, the LC-SFT group
% size $K$ and curvature suppression $\kappa$, the SCC-Verifier
% inference hyperparameters $M$, $N$, $T$, the text-only judge LLM
% used in the caption-as-evidence protocol, and the evaluation
% protocols for the four benchmarks will be finalised in this
% subsection.)}

\subsection{Caption-as-Evidence QA on Audio Benchmarks}
\label{sec:experiments:cae}

We adopt a \emph{caption-as-evidence} protocol: the captioner
generates a detailed audio caption for each clip, and a frozen
text-only judge language model (LLM-as-a-judge) answers the
benchmark question using only the caption as its source of
audio-derived evidence. The judge accuracy on
MMAU~\citep{mmau}, MMAR~\citep{mmar}, and MMSU~\citep{mmsu}
therefore measures the extent to which the caption stands in for
the audio itself: a higher score implies that the caption preserves
the fine-grained acoustic information required for downstream
reasoning, while a lower score indicates that audio facts are
either omitted or hallucinated and cannot be recovered from the
caption text. This protocol echoes the utility-based caption
evaluation paradigm advocated by \citet{captionqa}, in which a
frozen text-only LLM~\cite{qwen3.6-27b} answers benchmark questions using only the caption as its evidence source; high judge accuracy is taken to mean that the caption successfully \emph{stands in} for the underlying signal for downstream reasoning.

\paragraph{Training-evaluation consistency.}
This protocol is structurally identical to the training-time MCQ
reward of \S\ref{sec:method:reward}, differing only in the source
of the questions (gold-derived during training versus benchmark
MCQs at evaluation). The LC-SFT objective therefore directly
optimises the same caption-as-evidence quality that our evaluation
measures.

\begin{table}[t]
\centering
\small
\setlength{\tabcolsep}{4pt}
\begin{tabular}{l c c c}
\toprule
\textbf{Model} & \textbf{MMAU} & \textbf{MMAR} & \textbf{MMSU} \\
\midrule
\multicolumn{4}{l}{\textbf{\textit{Proprietary Models}}} \\
\midrule
GPT-4o Audio       & 62.4 & 59.3 & 56.4   \\
Gemini 2.0 Flash   & 58.6 & 50.6 & 51.0   \\
Gemini 2.5 Flash   & 65.6 & 58.2 & 58.1   \\
Gemini 2.5 Pro     & 70.0 & 64.1 & --     \\
\midrule
\multicolumn{4}{l}{\textbf{\textit{Open-Source Models}}} \\
\midrule
SALMONN-13B        & 58.4 & 42.5 & --     \\
MiDashengLM-7B     & 59.4 & 50.7 & --     \\
Qwen2-Audio-7B     & 63.3 & 44.2 & 53.3  \\
Qwen2.5-Omni-7B    & 65.2 & 51.8 & 60.6  \\
\midrule
\textbf{Ours}
                       & \textbf{68.3} & \textbf{57.0} & \textbf{62.4} \\
\bottomrule
\end{tabular}
\caption{\textbf{Caption-as-evidence QA accuracy} (\%) on MMAU,
MMAR, and MMSU. A frozen text-only judge LLM~\cite{qwen3.6-27b} answers benchmark questions using only the model-generated caption as evidence. The last row is our model.}
\label{tab:cae_audio_qa}
\end{table}

Our LC-SFT-trained Qwen2.5-Omni-7B improves over the vanilla
backbone from $65.2$ to $68.3$ on MMAU ($+3.1$), $51.8$ to $57.0$
on MMAR ($+5.2$), and $60.6$ to $62.4$ on MMSU ($+1.8$), attaining
the highest score among all open-source captioners on the three
benchmarks. It further surpasses GPT-4o Audio on MMAU ($68.3$ vs.\
$62.4$) and MMSU ($62.4$ vs.\ $56.4$), and narrows the gap to
Gemini 2.5 Pro to $1.7$ points on MMAU. The larger gain on MMAR,
which requires multi-step reasoning across mixed sound, music,
and speech, suggests that LC-SFT's intermediate-layer reweighting
most strongly benefits captions whose downstream reasoning depends
on transcript-faithful audio facts, a pattern reinforced by
MMSU's exclusive spoken-language focus.

\subsection{Detailed Captioning Evaluation on Omni-Cloze}
\label{sec:experiments:omnicloze}

The caption-as-evidence judge accuracy of
\S\ref{sec:experiments:cae} measures the downstream utility of a
caption but does not score the caption text itself. To complement
it with a direct caption-quality measurement, we evaluate on
\textbf{Omni-Cloze}~\citep{Omni-Captioner}, a cloze-style benchmark
that masks fine-grained spans in a gold caption and forces the
candidate captioner to fill them via multiple-choice selection,
with a ``not given'' option that distinguishes omission from
hallucination.
Table~\ref{tab:omni_cloze_audio_audio_only} reports accuracy on the
audio subset of Omni-Cloze under the audio-visual input setting, in
which the model receives both the video and audio streams. The
parallel audio-only input setting, where the model receives the
audio stream alone, is reported in
Appendix~\ref{app:omnicloze-audio-only}.

% \begin{table}[t]
% \centering
% \small
% \setlength{\tabcolsep}{6pt}
% \begin{tabular}{l c}
% \toprule
% \textbf{Model} & \textbf{Acc (\%) $\uparrow$} \\
% \midrule
% \multicolumn{2}{l}{\textbf{\textit{Proprietary Models}}} \\
% \midrule
% GPT-4o Audio       & 35.8 \\
% Gemini 2.0 Flash   & 20.0 \\
% Gemini 2.5 Flash   & 42.6 \\
% Gemini 2.5 Pro     & 48.0 \\
% Gemini3.1-Pro & \\
% Gemini3.1-Pro (with SCC) & \\
% \midrule
% \multicolumn{2}{l}{\textbf{\textit{Open-Source Models}}} \\
% \midrule
% SALMONN-13B            & 10.6 \\
% MiDashengLM-7B         & 19.5 \\
% Qwen2-Audio-7B         & 22.2 \\
% Qwen2.5-Omni-7B        & 25.8 \\
% \midrule
% \textbf{Ours}
%                        & \textbf{} \\
% \bottomrule
% \end{tabular}
% \caption{\textbf{Results on Omni-Cloze (audio-only split)}, audio-only input. The last row is our method based on Qwen2.5-Omni-7B.}
% \label{tab:omni_cloze_audio_audio_only}
% \end{table}

\begin{table}[t]
\centering
\small
\setlength{\tabcolsep}{6pt}
\begin{tabular}{l c}
\toprule
\textbf{Model} & \textbf{Accuracy (\%) $\uparrow$} \\
\midrule
\multicolumn{2}{l}{\textbf{\textit{Proprietary Models}}} \\
\midrule
Gemini 2.0 Flash                  &  31.7 \\
Gemini 2.5 Flash                  &  36.2 \\
Gemini 2.5 Pro                    &  44.1 \\
Gemini3.1-Pro            &  59.5 \\
Gemini3.1-Pro (with SCC) &  68.1 \\
\midrule
\multicolumn{2}{l}{\textbf{\textit{Open-Source Models}}} \\
\midrule
video-SALMONN-13B                 &   2.3 \\
videoLLaMA 2-7B                   &   4.3 \\
Qwen2.5-Omni-7B                   &  14.1 \\
video-SALMONN 2-7B                &  40.3 \\
Omni-Captioner-7B                 &  54.5 \\
\midrule
\textbf{Ours}
                       & \textbf{58.9} \\
\bottomrule
\end{tabular}
\caption{\textbf{Results on Omni-Cloze (audio-only split)}, audio-visual (video) input. The last row is our method based on Qwen2.5-Omni-7B.}
\label{tab:omni_cloze_audio_audio_only}
\end{table}

On the audio subset of Omni-Cloze under audio-visual input, our
LC-SFT-trained Qwen2.5-Omni-7B reaches $58.9\%$, a $+44.8$
absolute improvement over the vanilla backbone ($14.1\%$) and
$+4.4$ over the strongest open-source prior method
Omni-Captioner-7B~\citep{Omni-Captioner} ($54.5\%$). We surpass
Gemini 2.5 Pro by $14.8$ points and trail Gemini3.1-Pro
by only $0.6$. Applying the data and inference stages of SCC to
Gemini3.1-Pro yields a further $+8.6$ gain ($59.5\%$ to
$68.1\%$) without any training-stage modification, demonstrating
that SCC's self-check primitives function as model-agnostic
plug-in modules.

\subsection{Analysis}
\label{sec:experiments:analysis}

\paragraph{Self-check outcomes.}
Table~\ref{tab:gemini_self_check} reports all three evaluated settings on the
same human-validated Omni-Cloze evaluation set. Its Correct column provides
the accuracy summarised in Table~\ref{tab:omni_cloze_audio_audio_only}, while
Missing and Wrong expose the outcome decomposition. Net is Correct minus
Wrong, with Missing contributing zero. Adding the audio-only self-check to the
audio-visual draft yields the highest Correct rate, lowest Missing rate, and
best Net score, while Wrong rises from
$2.99\%$ to $4.66\%$.

\begin{table}[t]
\centering
\scriptsize
\caption{\textbf{Gemini3.1-Pro self-check outcomes.} Outcome rates (\%) on
the human-validated Omni-Cloze evaluation set. AV denotes audio-visual; Net is
Correct minus Wrong.}
\label{tab:gemini_self_check}
\setlength{\tabcolsep}{1.8pt}
\begin{tabular*}{\columnwidth}{@{\extracolsep{\fill}}lrrrr@{}}
\toprule
\textbf{Setting} & \textbf{Correct}$\uparrow$ & \textbf{Missing}$\downarrow$
 & \textbf{Wrong}$\downarrow$ & \textbf{Net}$\uparrow$ \\
\midrule
Audio-visual input              & 59.54 & 37.48 & \textbf{2.99} & 56.55 \\
Audio-only input                & 64.08 & 32.52 & 3.40 & 60.69 \\
AV + audio-only self-check      & \textbf{68.06} & \textbf{27.27}
                                & 4.66 & \textbf{63.40} \\
\bottomrule
\end{tabular*}
\end{table}

\paragraph{Verifier-format alignment.}
We test whether the LC-SFT improvement transfers beyond the MCQ/verifier
format and whether it depends on the text-judge family. First, SFT and LC-SFT
receive audio only and freely generate one caption for each clip in the complete
Clotho evaluation split~\citep{Clotho}, using the same prompt,
deterministic decoding, and a $512$-token limit. This control contains no
questions, answer options, caption-only judge, SCC-Verifier, or MCQ reranking.
The five metrics are Best- and Mean-Chunk Contrastive Language-Audio
Pretraining scores (B-CLAP and M-CLAP)~\citep{wu2023largeclap}, our
RefAlign-F1 (R-F1), Sentence-BERT similarity (SBERT)~\citep{SentenceBERT},
and Fluency-Enhanced Sentence-BERT Evaluation (FENSE)~\citep{FENSE}.
LC-SFT improves all five metrics over SFT, including B-CLAP from $0.460$ to
$0.492$ and FENSE from $0.401$ to $0.433$
(Table~\ref{tab:open_ended_clotho}).

We next freeze the Omni-Cloze audio-visual captions and replace only the
Qwen3.6-27B judge with Gemini3.1-Pro. LC-SFT remains above SFT, by
$+1.82$ points under Qwen and $+2.19$ under Gemini
(Table~\ref{tab:cross_model_judging}), so the gain transfers across judge
families.

\begin{table}[t]
\centering
\footnotesize
\caption{\textbf{Open-ended Clotho captioning.} Audio-only free generation on
the complete evaluation split. B-/M-CLAP denote best-/mean-chunk
Contrastive Language-Audio Pretraining; R-F1 denotes our RefAlign-F1; SBERT
denotes Sentence-BERT; and FENSE denotes Fluency-Enhanced Sentence-BERT Evaluation.
All metrics are higher-is-better.}
\label{tab:open_ended_clotho}
\setlength{\tabcolsep}{3.0pt}
\begin{tabular*}{\columnwidth}{@{\extracolsep{\fill}}lccccc@{}}
\toprule
\textbf{Model} & \textbf{B-CLAP} & \textbf{M-CLAP} & \textbf{R-F1}
 & \textbf{SBERT} & \textbf{FENSE} \\
\midrule
SFT    & 0.460 & 0.283 & 0.607 & 0.413 & 0.401 \\
LC-SFT & \textbf{0.492} & \textbf{0.311} & \textbf{0.612}
       & \textbf{0.436} & \textbf{0.433} \\
\bottomrule
\end{tabular*}
\end{table}

\begin{table}[t]
\centering
\footnotesize
\caption{\textbf{Cross-model judge robustness.} Omni-Cloze accuracy (\%) for
fixed audio-visual captions; only the text judge changes. Protocol details are
provided in Appendix~\ref{app:format-alignment-analysis}.}
\label{tab:cross_model_judging}
\begin{tabular*}{\columnwidth}{@{\extracolsep{\fill}}lcc@{}}
\toprule
\textbf{Captioner} & \textbf{Qwen3.6-27B} & \textbf{Gemini3.1-Pro} \\
\midrule
Qwen2.5-Omni-7B & 14.13 & 16.34 \\
SFT             & 57.10 & 59.51 \\
LC-SFT          & \textbf{58.92} & \textbf{61.70} \\
\bottomrule
\end{tabular*}
\end{table}

\paragraph{Recent baselines under a common protocol.}
Table~\ref{tab:recent_common_protocol} complements
Tables~\ref{tab:cae_audio_qa} and~\ref{tab:omni_cloze_audio_audio_only} with
three Qwen3-Omni variants~\citep{xu2025qwen3,Omni-Captioner}. Under the same
audio-only caption-as-evidence protocol, LC-SFT outperforms all three variants
on Omni-Cloze and MMAR, and is slightly higher on MMAU.

\begin{table}[t]
\centering
\scriptsize
\caption{\textbf{Comparison with Qwen3-Omni variants under a common audio-only
protocol.} Caption-as-evidence accuracy (\%). Bold marks the best result.}
\label{tab:recent_common_protocol}
\setlength{\tabcolsep}{2.4pt}
\renewcommand{\arraystretch}{0.94}
\begin{tabular*}{\columnwidth}{@{\extracolsep{\fill}}lccc@{}}
\toprule
\textbf{Model} & \textbf{Omni-Cloze} & \textbf{MMAR} & \textbf{MMAU} \\
\midrule
Qwen3-Omni-Captioner  & 57.5 & 55.3 & 66.4 \\
Qwen3-Omni-Instruct   & 35.3 & 54.5 & 67.9 \\
Qwen3-Omni-Thinking   & 33.6 & 52.6 & 63.8 \\
Ours                  & \textbf{59.3} & \textbf{57.0} & \textbf{68.3} \\
\bottomrule
\end{tabular*}
\end{table}

\subsection{Ablation of the SCC Framework}
\label{sec:experiments:ablation}

We decompose SCC into its three lifecycle stages and ablate each
contribution separately.

\paragraph{Data and inference stages.}
The last two proprietary rows of
Table~\ref{tab:omni_cloze_audio_audio_only} (Gemini3.1-Pro
with and without SCC) isolate the joint contribution of the data
and inference stages, applied as a wrapper on a frozen backbone.
SCC raises Omni-Cloze accuracy from $59.5\%$ to $68.1\%$
($+8.6$ absolute) without any training-stage modification,
confirming that these two stages function as model-agnostic
plug-in modules.

\paragraph{LC-SFT decomposition.}
We isolate the two reweighting components of LC-SFT in a
$2\!\times\!2$ design over the Stage-1 SFT baseline on the
Omni-Cloze audio subset: response-level advantage reweight alone
(\textsc{LC-SFT$_{\!R}$}, $A(y_k)$ from Eq.~\ref{eq:adv} with
$w_{k,t}\!\equiv\!1$), token-level layer-curvature reweight alone
(\textsc{LC-SFT$_{\!T}$}, $w_{k,t}$ from Eq.~\ref{eq:wt} with
$A(y_k)\!\equiv\!1$), and the full objective combining both.
Overall accuracy rises from $57.10$ (SFT) to $57.28$ with the
advantage signal alone ($+0.18$), to $58.59$ with the curvature
signal alone ($+1.49$), and to $58.92$ when both are combined
($+1.82$). The token-level curvature reweight is the dominant
contributor, winning on $17$ of $47$ categories; the advantage
signal is additive on top of curvature in $23$ categories,
especially speech- and narration-heavy ones. Full per-category
results are reported in Appendix~Table~\ref{tab:ablation_lc_sft}.

\section{Conclusion}
\label{sec:conclusion}

We introduced \textbf{Self-Check Captioning (SCC)}, an audio-grounded
question-answering framework for verifying long-paragraph detailed audio
captions during data construction, training, and inference.
\textbf{LACap-50k}, \textbf{LC-SFT}, and \textbf{SCC-Verifier} attain
state-of-the-art open-source results on MMAU, MMAR, MMSU, and Omni-Cloze.
Applied at the data and inference stages, SCC improves Gemini3.1-Pro by
$8.6$ Omni-Cloze points, demonstrating model-agnostic utility.

% ============================================================================
% Sec/limitations.tex --- Limitations section (required by EMNLP).
% Unnumbered (\section*) per EMNLP convention.
% ============================================================================

\section*{Limitations}
\label{sec:limitations}

LACap-50k is designed as a high-quality \emph{post-training}
corpus for long-paragraph detailed audio captioning, providing
$50{,}222$ carefully verified clips that suffice for supervised
fine-tuning and on-policy refinement on top of an already-capable
audio-visual backbone. Constructing a comparably high-quality
corpus at the scale required for \emph{pre-training} (typically
several orders of magnitude larger) remains prohibitively
expensive under our SCC pipeline, and we leave this extension to
future work.

\section*{Acknowledgments}
This work was supported by the CCF-Kuaishou Large Model Explorer Fund
(CCF-KuaiShou 2025002), the National Natural Science Foundation of China
(No.~62471420), and the GuangDong Basic and Applied Basic Research Foundation
(2025A1515012296).


\bibliography{custom}

\appendix

\section{LACap-50k Details}
\label{app:data}

\subsection{Audio Analysis Taxonomy}
\label{app:data:taxonomy}

A long-paragraph audio caption must traverse multiple, qualitatively
distinct facets of an audio signal from low-level acoustic properties
to high-level pragmatic intent. The seven-level taxonomy summarised in
Figure~\ref{fig:lacap:taxonomy} of the main text is encoded into both
the multimodal draft prompts and the audio-only relisten prompts of
the construction pipeline, so that every level the draft attempts to
describe is exactly a level the self-check pass probes. The levels
are:

\begin{itemize}
\item \textbf{Signal level}: recording quality, noise and artifacts,
acoustic space and reverberation, microphone distance, and other
properties intrinsic to the captured waveform.
\item \textbf{Perception level}: loudness and pitch, timbre and
rhythm, salience and spatial impression; how the audio is perceived
by a human listener.
\item \textbf{Semantic level}: directly verbalisable content,
including verbatim speech transcripts, song lyrics, identifiable
sound events, and the relative ordering of actions in time.
\item \textbf{Cultural level}: language and accent, genre tradition,
and the social or situational context implied by the audio.
\item \textbf{Pragmatic / intent / stance level}: communicative
purpose, emotional framing, and the speaker's attitude (explaining,
persuading, joking, instructing, warning, or evaluating).
\item \textbf{Musicological level}: tempo and rhythmic pattern,
instrumentation and vocal style, arrangement and production texture,
when music or musically structured audio is present.
\item \textbf{Foley / event sound level}: discrete non-speech event
sounds (impacts, handling sounds, interface tones, transitions, and
mechanical or scene cues), treated as first-class caption content
rather than collapsed into a vague ``background noise'' label.
\end{itemize}

Each level expands into three concrete subcategories (outer ring of
Figure~\ref{fig:lacap:taxonomy}); these subcategories are the units
at which the audio-only verifier formulates open-form relisten
questions during data construction.

\subsection{Lexical Evidence of Taxonomy Coverage}
\label{app:data:lexical}

To verify that the taxonomy is actually exercised by the released
captions (rather than only by the prompts), we aggregate the
vocabulary of all $50{,}222$ final captions into two facet-conditional
hot-word clouds shown in Figures~\ref{fig:lacap:wc_acoustic}
and~\ref{fig:lacap:wc_events} of the main text.
The \emph{acoustic-attribute} cloud is dominated by descriptors that
span the signal and perception levels of the taxonomy: dynamic and
spectral terms (\emph{loud}, \emph{faint}, \emph{bright},
\emph{high-pitched}, \emph{low-frequency}, \emph{muffled}),
articulation terms (\emph{sharp}, \emph{staccato}, \emph{sustained},
\emph{breathy}), recording-style terms (\emph{close-miked},
\emph{dry}, \emph{reverberant}, \emph{compressed}, \emph{distorted}),
and tempo-character terms (\emph{rhythmic}, \emph{syncopated},
\emph{upbeat}, \emph{fast-paced}, \emph{rapid}).
The \emph{sound-event} cloud is dominated by terms that span the
semantic and Foley levels: speech events (\emph{male speech},
\emph{female speech}, \emph{male voice speaking}, \emph{male
voiceover}), human-body events (\emph{intake of breath},
\emph{audible breaths}, \emph{lip smack}, \emph{footsteps}), ambient
and silence events (\emph{silence}, \emph{digital silence},
\emph{room tone}, \emph{background hiss}, \emph{broadband hiss}),
percussive musical events (\emph{kick drum}, \emph{hi-hats},
\emph{crash cymbal}, \emph{synthesized kick drum}), melodic events
(\emph{acoustic guitar strumming}, \emph{sub-bass}, \emph{bassline}),
and editorial events (\emph{abrupt cut to silence}, \emph{abrupt
audio cut}).

\subsection{Construction Pipeline Details}
\label{app:data:protocol}

The data-stage self-check of \S\ref{sec:data} is instantiated as a
sequence of independent inference passes using Gemini3.1-Pro-Preview at
temperature $0$, with no contextual state propagated across passes. The first pass receives the
complete original audio--video file, including clips longer than $120$ s,
without client-side sampling, segmentation, truncation, or additional
compression, and produces an initial draft caption. Each subsequent
verification pass is launched independently on the complete soundtrack,
extracted as a $16$-kHz mono WAV, and receives only that audio together with
one verification question generated from the draft. The video frames, the
draft caption, and earlier verification inputs and outputs are withheld. A
final aggregation pass uses the verification answers to revise the draft into
the released LACap-50k caption. The $0.2$ fps and $2$--$32$ frame setting in
\S\ref{sec:experiments:setup} applies only to Qwen2.5-Omni training, not to
Gemini data construction.

\subsection{Transcript-Fidelity Metric Definitions}
\label{app:data:asr}

Qwen3.5-27B reads only the final caption---not the audio, ASR transcript, or
error scores---and copies exact source spans containing explicit spoken, sung,
or chanted text while preserving their language and script. We exclude
translations, paraphrases, indirect speech, acoustic descriptions, and
non-human sound effects, and retain spans containing at least six Unicode
alphanumeric units (ten for short vocalisations). A caption is designated
speech-heavy when it contains at least $20$ retained units or at least two
valid spans, yielding $10{,}975$ captions and $36{,}809$ spans. Qwen3-ASR-1.7B
then transcribes each complete clip once. After Unicode normalisation, every
caption span is aligned to the minimum-Levenshtein local subsequence of the
full ASR transcript; residual paraphrased or indirect descriptions are
excluded rather than counted as ASR errors. The resulting values are therefore
an automatic consistency audit of explicit transcript spans, not a
human-ground-truth ASR evaluation or a full-caption factuality measure.

For each transcript span $s\!\in\!\mathcal{S}$ extracted from a
LACap-50k caption, let $h_s$ denote the hypothesis transcript taken
from the caption and $r_s$ the Qwen3-ASR reference. Let
$\mathrm{EditDist}_{c}(\cdot,\cdot)$ and
$\mathrm{EditDist}_{w}(\cdot,\cdot)$ denote the Levenshtein edit
distance at the character and word level, and let $|r_s|_{c},
|r_s|_{w}$ denote the reference length in characters and words. The
per-span error rates are
$\mathrm{CER}(s)\!=\!\mathrm{EditDist}_{c}(h_s,r_s)/|r_s|_{c}$ and
$\mathrm{WER}(s)\!=\!\mathrm{EditDist}_{w}(h_s,r_s)/|r_s|_{w}$.
We aggregate the character-level error rates across the span set
$\mathcal{S}$ in two complementary ways:
\begin{align}
\mathrm{Micro\text{-}CER}
  &= \frac{\sum_{s\in\mathcal{S}}\mathrm{EditDist}_{c}(h_s,r_s)}
          {\sum_{s\in\mathcal{S}}|r_s|_{c}},
   \label{eq:micro-cer} \\
\mathrm{Macro\text{-}CER}
  &= \frac{1}{|\mathcal{S}|}\sum_{s\in\mathcal{S}}\mathrm{CER}(s).
   \label{eq:macro-cer}
\end{align}
\emph{Micro} aggregates edit counts before normalising, so each span
contributes proportionally to its reference length and the
dataset-wide rate is dominated by long spans; \emph{macro} normalises
within each span first and then averages, treating short and long
spans equally. The word-level rates $\mathrm{Micro\text{-}WER}$ and
$\mathrm{Macro\text{-}WER}$ are defined analogously by replacing
character edit distance and character reference length with their
word-level counterparts. We additionally report the share of spans
whose per-span error rate stays below a conventional quality
threshold of $0.2$, namely
$\mathrm{CER}_{\le 0.2}\!=\!|\{s\!:\!\mathrm{CER}(s)\!\le\!0.2\}|/|\mathcal{S}|$
and the analogously defined $\mathrm{WER}_{\le 0.2}$; over LACap-50k,
$87.69\%$ of spans fall under the $0.2$ CER threshold and $82.55\%$
under the $0.2$ WER threshold. The macro rates ($8.32\%$ CER and
$11.51\%$ WER overall) sit slightly above the micro rates ($6.23\%$
and $9.21\%$), indicating a mild concentration of errors on shorter
spans rather than systematic long-span failure.

\subsection{Post-hoc Non-speech Event Grounding Audit}
\label{app:data:non-speech}

This audit complements the transcript-fidelity analysis by testing whether
explicit non-speech event claims in final captions are grounded in the source
audio.

\paragraph{Cohort and caption-only claim extraction.}
The formal cohort contains $1{,}000$ unique LACap-50k clips, stratified into
$500$ clips of $0$--$30$ s and $500$ clips of $30$--$60$ s. Clips used to
develop the extraction prompt, eligibility rules, or OpenFLAM thresholds are
excluded. At temperature $0$, Qwen3.6-27B~\citep{qwen3.6-27b} receives only
the final caption and extracts, for every explicit non-speech event, its source
text, a short detector-oriented query, and any stated time hint. All $1{,}000$
captions are parsed successfully.

\paragraph{Eligibility filtering.}
We exclude negated or absence claims,
ambiguous alternatives, compound queries without a single detection outcome,
silence and editing transitions, capture artifacts and vague background-noise
descriptions, extremely low-salience details, speculative causal
interpretations, and claims outside the detector's reliable semantic scope.
The remaining queries are canonicalised into short, self-contained event
phrases, and identical clip--query pairs are deduplicated. Retained claims span
human non-speech vocalisations, animals, vehicles and machinery, alarms and
electronics, impacts and household objects, natural sounds, movement, music
and percussion, crowds, and environmental ambience.

\paragraph{OpenFLAM inference.}
We use OpenFLAM~\citep{OpenFLAM}. Audio is loaded as a $48$-kHz mono waveform
and processed in overlapping $10$-s windows with a
$5$-s hop; short final windows are zero-padded. We run OpenFLAM in
\texttt{unbiased} frame-level mode, apply a width-$3$ median filter to each
similarity curve, and aggregate responses across all windows. Let $s_{\max}$
be the whole-clip peak response and $d_{\mathrm{active}}$ the accumulated
duration above the positive threshold. We label a query \texttt{corroborated}
when $s_{\max}\geq0.50$ and
$d_{\mathrm{active}}\geq0.20$ s, \texttt{not\_corroborated} when
$s_{\max}<0.35$, and \texttt{uncertain} otherwise.

\paragraph{Results.}
After eligibility filtering, canonicalisation, and deduplication, the cohort
contains $3{,}985$ claims: $3{,}770$ corroborated ($94.60\%$), $151$ not
corroborated ($3.79\%$), and $64$ uncertain ($1.61\%$). The expert-supported
fraction is $3770/3985=94.60\%$; decision coverage is
$(3770+151)/3985=98.39\%$; and support among decided claims is
$3770/(3770+151)=96.15\%$.

\section{Prompts}
\label{app:prompts}

This appendix collects the prompts used at two stages of the work:
(i)~the caption-to-MCQ construction that produces the training-time
reward of \S\ref{sec:method:reward}
(\S\ref{app:prompts:cap2mcq}); and (ii)~the cloze-to-MCQ
conversion that translates Omni-Cloze blanks into self-contained
five-option questions used in our evaluation pipeline
(\S\ref{app:prompts:cloze2mcq}). All prompt blocks below are
typeset verbatim from our released implementation;
\texttt{\{var\}} markers denote placeholders filled at call time.

\subsection{Caption-to-MCQ Construction (Training Reward)}
\label{app:prompts:cap2mcq}

At training time, the caption-only MCQ reward of
\S\ref{sec:method:reward} requires a pool $\mathcal{Q}_x$ of
five-option multiple-choice questions derived from the gold
caption $y_{\mathrm{gold}}$ of LACap-50k. The conversion is
performed offline by a frozen external text language model, once
per gold caption, before training begins; the prompt below
instructs the model to emit between 20 and 30 cloze-style MCQs
spanning the seven audio analysis levels of
\S\ref{app:data:taxonomy}, returning four model-generated options
A--D per question (the ``E: Not given'' option is appended by our
post-processor).

\begin{tcolorbox}[colback=gray!10,colframe=black,
    arc=1mm,auto outer arc,boxrule=0.5pt,breakable]
\footnotesize
\textbf{Caption-to-MCQ System Prompt}

\hrulefill

You are an expert audio comprehension MCQ generator. Given a
detailed audio caption, generate $20$--$30$ high-quality MCQs that
together cover all seven audio analysis levels of
\S\ref{app:data:taxonomy}. Each MCQ must test one specific,
clearly answerable audio detail, with multiple questions per level
wherever the caption provides evidence.

\textbf{Seven audio analysis levels.}
\begin{itemize}\setlength\itemsep{0.2em}
\item \texttt{signal\_level}: recording quality, mic technique,
reverberation, background noise, dynamic range, mixing balance,
stereo placement, artifacts.
\item \texttt{perception\_level}: loudness, pitch, timbre,
rhythmic feel, speech clarity, foreground vs.\ background
salience, perceived emotional intensity.
\item \texttt{semantic\_level}: exact words/phrases spoken
(ASR-style), identifiable sound events, speaker turns,
environmental sounds, relative ordering of audio events.
\item \texttt{cultural\_level}: speaker language and accent,
musical genre tradition, social or situational context.
\item \texttt{pragmatic\_level}: communicative purpose,
rhetorical style, speaker stance, whether the voice is human or
synthetic.
\item \texttt{musicological\_level}: tempo, instrumentation,
vocal style, music structure, production texture, genre.
\item \texttt{sound\_effect\_level}: specific sound effect types,
discrete event sounds, Foley-like action sounds.
\end{itemize}

\textbf{Rules.}
\begin{itemize}\setlength\itemsep{0.2em}
\item Every question must be answerable solely by listening to
the audio; the correct answer must be directly supported by the
caption text.
\item Generate four choices A--D only; the post-processor
appends ``E: Not given'' automatically. Distractors must be
realistic within the same category.
\item Order-of-event questions must use relative terms
(``first'', ``after'', ``last'') and must not mention specific
timestamp numbers.
\item Each question is concise (max $25$ words);
\texttt{answer\_text} must exactly match the chosen option.
\item Output ONLY a valid JSON array; no markdown, no code
fences, no reasoning.
\end{itemize}

\textbf{Output format (per MCQ object).}
\begin{verbatim}
{"level": "<level>",
 "subcategory": "<aspect>",
 "question": "<text>",
 "A": "<text>", "B": "<text>",
 "C": "<text>", "D": "<text>",
 "answer": "<A|B|C|D>",
 "answer_text": "<matches option>"}
\end{verbatim}
\end{tcolorbox}

\subsection{Omni-Cloze to MCQ Conversion (Evaluation Preparation in Fig.~\ref{fig:gen_mode_failure:left})}
\label{app:prompts:cloze2mcq}

The Omni-Cloze benchmark~\citep{Omni-Captioner} presents
captions as cloze passages with placeholders such as
\texttt{[BLANK\_1]}, \texttt{[BLANK\_2]}, \dots, each paired
with five candidate options. To evaluate our LC-SFT-trained
captioner against Omni-Cloze under the caption-as-evidence
protocol of \S\ref{sec:experiments:cae}, we convert each blank
into a self-contained natural-language question that can be
answered by reading the caption text alone, while preventing the
question from leaking answers to neighbouring blanks. The
conversion is implemented as two complementary prompts: a
per-blank prompt (used as the default and as the fallback for any
sample-level failure), and a sample-level batched prompt (used
for efficiency when many blanks share a passage). Visual-only
blanks have their answers pre-substituted into the passage so
that the LLM sees a fully grammatical context; only MCQ-relevant
blanks (audio and audio-visual) remain as unknown
\texttt{[BLANK\_M]} markers.

\begin{tcolorbox}[colback=gray!10,colframe=black,
    arc=1mm,auto outer arc,boxrule=0.5pt,breakable]
\footnotesize
\textbf{Cloze-to-MCQ System Prompt (Per-Blank, Anti-Leakage)}

\hrulefill

You are an expert at creating video comprehension test questions.
The passage contains blanks written as
\texttt{[BLANK\_1]}, \texttt{[BLANK\_2]}, \dots,
\texttt{[BLANK\_30]}. Your task: write ONE concise question
whose answer fills the TARGET blank specified in the prompt;
every other \texttt{[BLANK\_M]} is a completely unknown value.

\textbf{Rules.}
\begin{itemize}\setlength\itemsep{0.2em}
\item Output ONLY the question text (no numbering, no
explanation, no answer); one sentence, max $25$ words.
\item Do NOT state or imply the answer to the target blank in
the question.
\item Do NOT guess, infer, or mention specific values for any
other \texttt{[BLANK\_M]}: every other blank is unknown, never
write a word you think fills it.
\item Frame the question as something a viewer answers from the
video using the modality hint provided.
\end{itemize}
\end{tcolorbox}

\begin{tcolorbox}[colback=gray!10,colframe=black,
    arc=1mm,auto outer arc,boxrule=0.5pt,breakable]
\footnotesize
\textbf{Cloze-to-MCQ System Prompt (Sample-Level Batched)}

\hrulefill

You are an expert at creating video comprehension test questions.
The passage contains blanks written as \texttt{[BLANK\_M]}. Your
task: write concise questions for the specified TARGET blanks
only; every \texttt{[BLANK\_M]} marker is an unknown value unless
it appears as plain text.

\textbf{Rules.}
\begin{itemize}\setlength\itemsep{0.2em}
\item Output ONLY valid JSON whose keys are blank numbers as
strings; no markdown, no explanation.
\item Each value is one natural question, one sentence, max
$25$ words.
\item Do NOT state or imply the answer to a target blank in its
own question.
\item Do NOT guess, infer, or mention specific values for any
other \texttt{[BLANK\_M]}.
\item Frame each question as something a viewer answers from
the video using the modality hint.
\end{itemize}

\textbf{Output format.}
\begin{verbatim}
{"3": "<question for blank 3>",
 "7": "<question for blank 7>"}
\end{verbatim}
\end{tcolorbox}

\paragraph{Two-layer anti-leakage protocol.}
The two prompts above are combined with a post-generation
validation pass: for each generated question, we check whether
any other MCQ blank's answer string appears verbatim (whole-word,
case-insensitive); if so, the question is regenerated with an
explicit forbidden-word constraint appended to the prompt, up to
a fixed retry budget. Unresolved items are flagged with
\texttt{\_leakage\_warning} for downstream filtering. Non-MCQ
(visual-only) blanks have their answers pre-substituted into the
passage rather than masked, since the LLM never sees a
\texttt{[BLANK\_M]} marker that it is not authorized to discuss;
this also gives the LLM a fully grammatical reading context
instead of a sentence riddled with placeholders.

\section{Method Details}
\label{app:method}

\subsection{Why Second-Order Curvature}
\label{app:method:curv-design}

We define $\mathrm{Curv}_{k,t}$ in Eq.~\ref{eq:curv} as the smoothed
second-order difference of $U^{(\ell)}_{k,t}$ across late layers,
rather than the first-order alternative
$\sum_{j}\!|\Delta U^{(j)}_{k,t}|$. The choice is essential to
isolating SEC. The second-order formulation evaluates to zero on any
constant-slope trajectory (stable, linearly ascending, or linearly
descending), and is large only on trajectories that exhibit a true
kink between consecutive intermediate layers. The first-order
alternative cannot make this distinction. It fires equally on a
smoothly resolving common token, where each $\Delta U^{(j)}_{k,t}$
is small but nonzero, and on an abrupt late-layer commitment, where
$\Delta U^{(j)}_{k,t}$ spikes at one position. Only the second-order
statistic produces near-zero weight on smooth descents and
concentrates suppression on the kinked trajectories that diagnose
SEC.

\subsection{Full Gradient Expression}
\label{app:method:gradient}

Differentiating the group-level objective in Eq.~\ref{eq:groupobj}
with respect to $\theta$ yields
\begin{equation}
\nabla_\theta\mathcal{L}_{\mathrm{LC}}
\;=\;
-\,\mathbb{E}_{x,\mathcal{G}_x}\!\!\left[
\sum_{k=1}^{K}\frac{A(y_k)}{Z_k}
\sum_{t=1}^{T_k} w_{k,t}\,\nabla_\theta\ell_{k,t}
\right],
\label{eq:gradinterp}
\end{equation}
where $Z_k\!=\!\sum_t w_{k,t}\!+\!\epsilon$ is the per-candidate
normalizer. The expression makes the two-axis reweighting explicit.
The candidate axis is reweighted by $A(y_k)/Z_k$, which is positive
for above-mean candidates (maximum-likelihood gradient pulling
$\pi_\theta$ towards them) and negative for below-mean candidates
(unlikelihood gradient pushing $\pi_\theta$ away). Within each
candidate the token axis is reweighted by $w_{k,t}$, close to one
for smoothly-resolved tokens and close to zero for SEC-prone tokens.
The combined effect concentrates gradient mass simultaneously on
high-reward candidates and on intermediate-layer-reliable tokens.

\subsection{Two-Stage Training Details}
\label{app:method:two-stage}

Stage~1 applies standard cross-entropy SFT on
$\mathcal{D}\!=\!\{(x,y^{\star})\}$ where $y^{\star}$ is the
LACap-50k gold caption for $x$. The model from this stage gains the
basic capacity to produce paragraph-length captions for
audio--visual clips. Stage~2 applies LC-SFT as defined in
\S\ref{sec:method}, initialized from the Stage~1 checkpoint. All
rollouts $y_{r_1},\dots,y_{r_{K-1}}$ in the response group
$\mathcal{G}_x$ are sampled from the current LC-SFT policy
$\pi_\theta$ at every gradient step. The Stage~1 checkpoint is used
only for initialization and is not consulted during Stage~2. The
optimizer, learning rate, batch size, group size $K$, curvature
suppression coefficient $\kappa$, GPU configuration, and total
training steps for both stages are reported in
\S\ref{sec:experiments}.

\subsection{Relation to the Training Reward}
\label{sec:inference:vs-training}

SCC-Verifier and the caption-only MCQ reward of
\S\ref{sec:method:reward} share the same verification primitive
(audio-grounded multiple-choice question answering) but differ in
three respects, each dictated by the constraints of the lifecycle
stage they operate at:

\begin{itemize}
\item \textbf{Source of MCQs.}
At training, MCQs are derived from the data-stage SCC-verified gold
caption $y_{\mathrm{gold}}$; at inference, no gold caption is
available, so MCQs are derived from each candidate $y^{(m)}$ itself.

\item \textbf{Verifier identity.}
At training, the verifier is a frozen external text language model
that scores caption-text retention without listening to the audio,
forcing the candidate caption to preserve gold-verified facts in
its text. At inference, the verifier is the LC-SFT-trained model
itself listening to the audio without seeing the candidate caption,
forcing the candidate's own claims to be audio-grounded.

\item \textbf{Role in the SCC framework.}
Together with the data-stage modality-subtraction self-check
(\S\ref{sec:data:stats}), the two complete SCC's audio-grounded
verification primitive across all three lifecycle stages of a
long-paragraph audio caption.
\end{itemize}

\section{Omni-Cloze Audio-Only Input Results}
\label{app:omnicloze-audio-only}

Each Omni-Cloze clip provides both an audio stream and a video
stream. The main-text Table~\ref{tab:omni_cloze_audio_audio_only}
reports accuracy on the audio subset of Omni-Cloze under the
audio-visual input setting, in which the model receives both
modalities. For completeness, we additionally evaluate the same
audio subset under the audio-only input setting, in which the model
receives the audio stream alone with no access to the video. This
complementary setting isolates each model's audio-understanding
capability from any visual disambiguation that the video stream
might provide. Table~\ref{tab:omni_cloze_audio_only_appendix}
reports the results.

\begin{table}[t]
\centering
\small
\setlength{\tabcolsep}{6pt}
\begin{tabular}{l c}
\toprule
\textbf{Model} & \textbf{Acc (\%) $\uparrow$} \\
\midrule
\multicolumn{2}{l}{\textbf{\textit{Proprietary Models}}} \\
\midrule
GPT-4o Audio~\citep{gpt4o}                  & 35.8 \\
Gemini 2.0 Flash~\citep{gemini}             & 20.0 \\
Gemini 2.5 Flash~\citep{gemini}             & 42.6 \\
Gemini 2.5 Pro~\citep{gemini}               & 48.0 \\
Gemini3.1-Pro~\citep{gemini}                & 64.1 \\
\midrule
\multicolumn{2}{l}{\textbf{\textit{Open-Source Models}}} \\
\midrule
SALMONN-13B~\citep{Salmonn}                 & 10.6 \\
MiDashengLM-7B~\citep{Midashenglm}          & 19.5 \\
Qwen2-Audio-7B~\citep{Qwen2Audio}           & 22.2 \\
Qwen2.5-Omni-7B~\citep{qwen25omni}          & 25.8 \\
Audio-Captioner-7B~\citep{Omni-Captioner}   & 53.2 \\
\midrule
\textbf{Qwen2.5-Omni-7B (Ours)}
                       & \textbf{59.3} \\
\bottomrule
\end{tabular}
\caption{\textbf{Results on Omni-Cloze (audio subset) under the
audio-only input setting.} Each model receives only the audio
stream, with no access to the video. The last row is our
LC-SFT-trained model based on Qwen2.5-Omni-7B.}
\label{tab:omni_cloze_audio_only_appendix}
\end{table}

\section{LC-SFT Component Ablation}
\label{app:ablation-lc-sft}

We isolate the two reweighting components of the LC-SFT objective
(\S\ref{sec:method}) in a $2\!\times\!2$ design on top of the
Stage-1 SFT baseline. \textbf{SFT} is the Stage-1 standard SFT
checkpoint on LACap-50k. \textbf{+Adv} applies only the
response-level advantage reweight $A(y_k)$ from Eq.~\ref{eq:adv}
while keeping $w_{k,t}\!\equiv\!1$ (denoted \textsc{LC-SFT$_{\!R}$}).
\textbf{+Tok} applies only the token-level layer-curvature reweight
$w_{k,t}$ from Eq.~\ref{eq:wt} while keeping $A(y_k)\!\equiv\!1$
(denoted \textsc{LC-SFT$_{\!T}$}). \textbf{LC-SFT} is the full
objective combining both. All four variants are evaluated on the
audio subset of Omni-Cloze under audio-visual input.
Table~\ref{tab:ablation_lc_sft} reports per-category correct
accuracy (\%) on all $47$ Omni-Cloze audio-analysis categories,
together with the overall accuracy aggregated across the full split.

\begin{table*}[t]
\centering
\footnotesize
\setlength{\tabcolsep}{3pt}
\renewcommand{\arraystretch}{1.02}
\begin{tabular}{@{}l rrrr @{\hspace{14pt}} l rrrr@{}}
\toprule
\textbf{Category} & \textbf{SFT} & \textbf{+Adv} & \textbf{+Tok} & \textbf{LC-SFT} &
\textbf{Category} & \textbf{SFT} & \textbf{+Adv} & \textbf{+Tok} & \textbf{LC-SFT} \\
\midrule
Artificial Intelligence  & 59.41 & 61.19 & \textbf{61.98} & 60.20 & Lectures \& Talks        & 57.23 & 58.42 & \textbf{61.58} & 60.40 \\
Arts \& Crafts           & 65.00 & 63.00 & \textbf{67.20} & 66.20 & Listicles \& Rankings    & 59.80 & \textbf{61.98} & 60.20 & 60.79 \\
Athlete Profiles         & 50.92 & 51.74 & 50.51 & \textbf{53.99} & Livestreams              & \textbf{56.77} & 52.62 & 55.90 & 56.55 \\
Board Games \& Puzzles   & \textbf{58.20} & 56.80 & 56.20 & 55.40 & Miscellaneous            & 57.38 & 53.48 & 55.94 & \textbf{58.81} \\
Business \& Finance      & 58.45 & 59.67 & \textbf{59.88} & 57.43 & Motorsports              & 52.68 & 55.27 & 57.46 & \textbf{58.45} \\
Car Modifications        & 62.00 & 57.09 & \textbf{62.38} & 61.06 & Movies \& Trailers       & 51.59 & \textbf{51.98} & 50.60 & 50.99 \\
Car Reviews              & 59.51 & 59.51 & \textbf{63.97} & 57.49 & Music                    & 40.84 & 40.08 & 40.27 & \textbf{41.60} \\
Collecting               & 61.13 & 60.53 & \textbf{63.56} & 61.94 & News Reports             & 57.74 & 59.52 & 59.52 & \textbf{62.30} \\
Comedy                   & 49.60 & \textbf{51.00} & 49.00 & 49.40 & Off-Roading              & 52.26 & 51.32 & \textbf{53.01} & 52.44 \\
Crafts \& DIY            & 60.04 & 58.56 & \textbf{62.16} & 60.47 & Opinion \& Commentary    & 56.47 & 58.43 & 57.84 & \textbf{59.41} \\
Debates                  & \textbf{56.50} & 55.92 & 55.92 & 55.92 & Outdoor Activities       & 57.51 & 57.14 & 58.44 & \textbf{60.85} \\
Driving Tutorials        & 58.88 & 61.57 & 63.84 & \textbf{64.67} & Parenting \& Family      & 56.75 & 60.76 & \textbf{61.60} & 59.28 \\
Engineering              & 67.57 & 64.64 & \textbf{67.78} & \textbf{67.78} & Photography \& Film      & 51.47 & 52.05 & 51.47 & \textbf{52.84} \\
Environmental Science    & 57.79 & 57.99 & \textbf{61.07} & 59.84 & Science \& Technology    & \textbf{67.36} & 66.32 & 67.15 & 66.74 \\
Fashion \& Beauty        & 60.51 & 59.72 & 61.89 & \textbf{63.26} & Science Explainers       & 55.76 & 57.41 & 59.88 & \textbf{60.29} \\
Fitness \& Training      & 63.95 & 64.77 & 63.95 & \textbf{69.45} & Space Exploration        & 62.98 & 64.89 & \textbf{65.46} & 63.93 \\
Food \& Cooking          & 60.86 & 60.45 & 61.27 & \textbf{62.09} & Sports Commentary        & 57.36 & \textbf{62.05} & 61.19 & 61.62 \\
Gaming                   & 53.41 & 56.63 & \textbf{56.83} & 56.02 & Tech Reviews             & 59.27 & 58.25 & 59.06 & \textbf{60.49} \\
Health \& Fitness        & 55.98 & 52.39 & 54.78 & \textbf{57.97} & Travel                   & 57.61 & 61.32 & \textbf{63.37} & 60.08 \\
Highlights \& Replays    & 49.41 & 50.00 & 51.19 & \textbf{51.78} & Tutorials                & 59.88 & 59.46 & \textbf{62.58} & 62.16 \\
History \& Culture       & 57.03 & 53.66 & 55.45 & \textbf{58.02} & Visual Arts              & 56.14 & 57.95 & 59.56 & \textbf{59.76} \\
Home \& Garden           & 60.29 & 61.51 & 62.73 & \textbf{63.34} & Vlogs                    & \textbf{53.58} & 50.10 & 50.10 & 50.87 \\
Interviews               & 53.60 & 53.81 & 57.63 & \textbf{59.53} & Writing \& Literature    & 54.09 & 55.58 & 54.99 & \textbf{57.73} \\
Language Learning        & 53.35 & 55.91 & 58.27 & \textbf{60.24} &                          &       &       &       &       \\
\midrule
\multicolumn{10}{c}{\textbf{Overall accuracy:}\quad SFT $57.10$\quad +Adv $57.28$\quad +Tok $58.59$\quad \textbf{LC-SFT $58.92$}} \\
\bottomrule
\end{tabular}
\caption{\textbf{LC-SFT component ablation on Omni-Cloze} (audio
subset, audio-visual input). Per-category correct accuracy (\%) on
all $47$ audio-analysis categories for the four variants on top of
Qwen2.5-Omni-7B. \textbf{SFT}: Stage-1 standard SFT on LACap-50k.
\textbf{+Adv}: Stage-2 with response-level advantage reweight only
($A(y_k)$ from Eq.~\ref{eq:adv}, $w_{k,t}\!\equiv\!1$; denoted
\textsc{LC-SFT$_{\!R}$}). \textbf{+Tok}: Stage-2 with token-level
layer-curvature reweight only ($w_{k,t}$ from Eq.~\ref{eq:wt},
$A(y_k)\!\equiv\!1$; denoted \textsc{LC-SFT$_{\!T}$}).
\textbf{LC-SFT}: full objective combining both reweighting signals.
Per-row best is in bold; ties are bolded jointly.}
\label{tab:ablation_lc_sft}
\end{table*}

\section{Relation between SCC-Verifier and the Training Reward}
\label{app:inference:relation-to-training}

SCC-Verifier (\S\ref{sec:inference}) and the caption-only MCQ
reward of \S\ref{sec:method:reward} share the same verification
primitive, namely audio-grounded multiple-choice question
answering, but operate at different lifecycle stages with
complementary rather than conflicting roles.

At training time, MCQs are derived from the data-stage SCC-verified
gold caption $y_{\mathrm{gold}}$, and the verifier is a frozen
external text language model that scores the candidate caption
$y_k$ on its text alone with no audio access. The reward thereby
forces $y_k$ to retain every audio fact already verified in
$y_{\mathrm{gold}}$, turning the training signal into a measure of
verified audio-fact retention from the gold reference. The
text-only verifier identity is deliberate. Allowing the training
verifier to access the audio would let it bypass $y_k$ and answer
from perception alone, leaving the candidate text unused and the
gradient uninformed about which facts the candidate actually
retains.

At inference time, no gold caption is available, so MCQs are
derived from each candidate $y^{(m)}$ itself. The verifier is the
LC-SFT-trained model, queried in audio-only MCQ mode with the
candidate's text deliberately withheld. The score thereby forces
each candidate's own assertions to be audio-grounded, with the
trained model acting as a self-consistency oracle on the audio
side. The audio-grounded verifier identity is again deliberate.
Allowing the inference verifier to read the candidate caption
would collapse the check into textual self-agreement and would not
test whether the assertions correspond to the actual audio.

The two mechanisms are therefore complementary by construction.
The training reward exploits the data-stage SCC verification (the
gold caption is already audio-checked) to teach the policy what
audio-grounded captioning looks like, using a text-only judge to
keep the supervision focused on caption-side fact retention. The
inference verifier exploits the resulting capability of the trained
model itself, using audio-grounded self-answering to arbitrate
among samples without any gold reference. Together with the
data-stage modality-subtraction self-check
(\S\ref{sec:data:stats}), the two complete SCC's audio-grounded
verification primitive across all three lifecycle stages of a
long-paragraph audio caption, with the verifier's modality tuned
to what each stage requires.

\section{Experiment Setup}
\label{sec:experiments:setup}

\paragraph{Backbone and training framework.}
We adopt Qwen2.5-Omni-7B~\citep{qwen25omni} as the captioner
backbone for both stages of training. Training is implemented with
the Swift framework using DeepSpeed ZeRO-3 on $8$ GPUs with
\texttt{bf16} mixed precision and gradient checkpointing. The
random seed is fixed at $42$.

\paragraph{Two-stage training and data split.}
The $50{,}222$-clip LACap-50k corpus is partitioned into two
disjoint training subsets. Stage~1 (standard SFT) uses $44$k clips
for cross-entropy supervised fine-tuning on the gold long-paragraph
captions, while Stage~2 (LC-SFT) uses a separate $1$k clips for
on-policy refinement under the caption-only MCQ reward and the
layer-curvature reweighting signal (\S\ref{sec:method}). The two
stages share the same optimisation configuration. Both use
learning rate $1\!\times\!10^{-6}$ with cosine schedule and warmup
ratio $0.03$, per-device batch size $1$, gradient accumulation $2$,
$1$ training epoch, and a maximum sequence length of $32{,}768$
tokens with the delete truncation strategy. The only differences
between the two stages are the training objective (cross-entropy
in Stage~1 versus the LC-SFT loss of Eq.~\ref{eq:groupobj} in
Stage~2) and the respective data subset.

\paragraph{Multimodal input processing.}
Each clip is processed jointly as a synchronised video and audio
stream, with the audio output head disabled so that the model
produces text only. Video frames are sampled at $0.2$ frames per
second within a range of $2$ to $32$ frames per clip. The video
token budget per clip is $64$ to $128$ tokens, and the image token
budget is $4$ to $512$ tokens.

\paragraph{LC-SFT and SCC-Verifier hyperparameters.}
For Stage~2 LC-SFT, the response group size is $K\!=\!4$ (one
LACap-50k gold caption plus three on-policy rollouts), the
curvature suppression coefficient is $\kappa\!=\!2$, and the
late-layer set $\mathcal{L}$ comprises the last $m\!=\!3$ Thinker
layers. For inference-time SCC-Verifier, we use $M\!=\!4$
candidate captions (one greedy plus three multinomial samples
at temperature $T\!=\!0.8$, symmetric with the training-time group
size $K$) and $N$ verification MCQs per candidate. We do not fix $N$ but instead extract every atomic
audio-grounded claim that the external LLM can identify in the
candidate caption, so longer captions naturally yield more
verification MCQs and a more fine-grained per-candidate score.

\paragraph{Evaluation setup.}
For the caption-as-evidence protocol of
\S\ref{sec:experiments:cae}, the frozen text-only judge LLM is
Qwen3.6-27B~\citep{qwen3.6-27b}. Prior-method numbers on MMAU and
MMAR are reproduced from~\citet{Omni-Captioner}.

\paragraph{Baselines.}
We compare against representative proprietary and open-source
audio-visual captioners across MMAU
(Table~\ref{tab:cae_audio_qa}), MMAR (Table~\ref{tab:cae_audio_qa}),
MMSU (Table~\ref{tab:cae_audio_qa}), and Omni-Cloze
(Table~\ref{tab:omni_cloze_audio_audio_only} for audio-visual input
and Table~\ref{tab:omni_cloze_audio_only_appendix} for audio-only
input). The proprietary baselines are
GPT-4o Audio~\citep{gpt4o} and the Gemini family of multimodal
models~\citep{gemini}, evaluated at the 2.0~Flash, 2.5~Flash,
2.5~Pro and 3.1~Pro-Preview release tiers. The open-source
baselines on audio-only inputs are SALMONN-13B~\citep{Salmonn},
MiDashengLM-7B~\citep{Midashenglm},
Qwen2-Audio-7B~\citep{Qwen2Audio}, and
Qwen2.5-Omni-7B~\citep{qwen25omni}, the last of which also serves
as our backbone. The open-source baselines on audio-visual inputs
additionally include the video-grounded captioners
video-SALMONN-13B~\citep{VideoSalmonn},
videoLLaMA~2-7B~\citep{video-llama},
video-SALMONN~2-7B~\citep{VideoSALMONN2}, and the long-paragraph
detailed captioner Omni-Captioner-7B~\citep{Omni-Captioner} (also
denoted Audio-Captioner-7B in the audio-only setting).

\section{Analysis Protocols Beyond Verifier-Format Alignment}
\label{app:format-alignment-analysis}

\paragraph{Open-ended Clotho evaluation.}
SFT and LC-SFT receive audio only and generate one caption for every clip in
the complete $1{,}045$-clip Clotho evaluation split~\citep{Clotho}, using the
same captioning prompt, deterministic decoding, and a $512$-token limit.
Generation and evaluation contain no questions, answer options, caption-only
judge, SCC-Verifier, or MCQ reranking. Because the generated paragraphs are
much longer than Clotho's five short references, Best- and Mean-Chunk
CLAP~\citep{wu2023largeclap} score consecutive semantic chunks of at most $30$
words.

\paragraph{RefAlign-F1 (ours).}
We split each generated caption into ordered sentence or clause units
$C=\{c_i\}_{i=1}^{m}$ and denote its five references by
$R=\{r_j\}_{j=1}^{n}$, where $n=5$. Splitting follows punctuation and line
breaks; overlong units are further divided into consecutive word chunks. We
encode all units and references with BGE-M3~\citep{BGE-M3}, $L_2$-normalise the
embeddings, and define $S_{ij}=\max\{0,\cos(e(c_i),e(r_j))\}$. With word-count
weights $w_i=\max\{1,\operatorname{words}(c_i)\}/\sum_k
\max\{1,\operatorname{words}(c_k)\}$, alignment precision and recall are
$P=\sum_i w_i\max_j S_{ij}$ and $R=n^{-1}\sum_j\max_i S_{ij}$. RefAlign-F1 is
$2PR/(P+R)$ when $P+R>0$ and zero otherwise; we report the arithmetic mean of
per-clip scores. This score measures soft alignment to the five references;
audio grounding is evaluated separately. For SBERT and FENSE,
BGE-M3 selects the complete generated sentence with the highest mean reference
similarity before scoring.

\paragraph{Cross-family judge control.}
All captioners first generate captions from synchronised audio--visual input.
We freeze these captions, after which Qwen3.6-27B and Gemini3.1-Pro
each receive the identical caption, question, and answer options without
audio/video access. Judge prompts, temperature ($0$), and answer parsing are
fixed, so the only changed variable is judge identity.

% \section{Example Appendix}
% \label{sec:appendix}

% This is an appendix.

\end{document}